\documentclass[twocolumn,pre,floatfix,superscriptaddress]{revtex4} % ,nofootinbib

\usepackage{dcolumn}
\usepackage{graphicx}
\usepackage{rotating}
\usepackage[usenames,dvipsnames]{color}
\usepackage{multirow}
\usepackage{amsmath,amsfonts,amssymb}
\usepackage{todonotes}
\usepackage{hyperref}

\newcommand{\remove}[1]{}

\begin{document}
%\onehalfspacing

\title{Escalation dynamics and the severity of wars}

\author{Aaron Clauset}
\email{aaron.clauset@colorado.edu}
\affiliation{Department of Computer Science, University of Colorado, Boulder, CO 80309, USA}
\affiliation{BioFrontiers Institute, University of Colorado, Boulder, CO 80303, USA}
\affiliation{Santa Fe Institute, Santa Fe, NM 87501, USA}
\author{Barbara F. Walter}
\affiliation{School of Global Policy and Strategy, University of California, San Diego, CA 92093, USA}
\author{Lars-Erik Cederman}
\affiliation{Center for Comparative and International Studies, ETH Z\"urich, 8092 Z\"urich, Switzerland}
\author{Kristian Skrede Gleditsch}
\affiliation{Department of Government, University of Essex, CO7 9QF UK}
\affiliation{Peace Research Institute Oslo, 0186 Oslo, Norway}

%\date{\today}

\begin{abstract} % 226 < 250 words
Although very large wars remain an enduring threat in global politics, we lack a clear understanding of how some wars become large and costly, while most do not. There are three possibilities:\ large conflicts start with and maintain intense fighting, they persist over a long duration, or they escalate in intensity over time. Using detailed within-conflict data on civil and interstate wars 1946--2008, we show that escalation dynamics---variations in fighting intensity within an armed conflict---play a fundamental role in producing large conflicts and are a generic feature of both civil and interstate wars. However, civil wars tend to deescalate when they become very large, limiting their overall severity, while interstate wars exhibit a persistent risk of continual escalation. A non-parametric model demonstrates that this distinction in escalation dynamics can explain the differences in the historical sizes of civil vs.\ interstate wars, and explain Richardson's Law governing the frequency and severity of interstate conflicts over the past 200 years. Escalation dynamics also drive enormous uncertainty in forecasting the eventual sizes of both hypothetical and ongoing civil wars, indicating a need to better understand the causes of escalation and deescalation within conflicts. The close relationship between the size, and hence the cost, of an armed conflict and its potential for escalation has broad implications for theories of conflict onset or termination and for risk assessment in international relations.
\end{abstract}

\maketitle

% 117 < 120 words
\noindent \textbf{Significance}: 
Wars are an enduring threat worldwide to social and economic stability. However, existing theories say little about how the most destructive wars become so large in the first place. Analyzing disaggregated data on the annual severities of civil and interstate wars worldwide since 1946, we show that large conflicts are caused by escalation, in which fighting intensity grows over time. A simple model of conflict escalation dynamics is sufficient to explain the full variation in historical sizes of both civil and interstate wars, while models without escalation fail to produce sufficiently large conflicts. Escalation dynamics are thus a fundamental feature of armed conflict, and provide a rich new analytic window to understanding war deterrence, onset, and termination.

\bigskip
\noindent \textbf{Introduction}
% 1. 
War remains a persistent feature of global politics, and the risk of a very large war is an enduring threat worldwide~\cite{gleditsch:clauset:2018}. At the same time, we lack a general understanding of how some wars become so large and highly destructive, while others do not. Do large wars start with very intense fighting at the outset? Or, does a conflict become large by simply lasting longer, providing more time to accumulate casualties? Or, does a war become large by escalating to more intense fighting? Does fighting intensity within a conflict unfold in characteristic ways, e.g., tending to escalate or tending to deescalate over time? In an ongoing conflict, how likely is an escalation or a deescalation in fighting over the next year, and by how much? A deeper understanding of how the sizes of armed conflicts accumulate over their durations would shed new light on the mechanisms that generate large wars, inform theories of war onset and risk assessment in international relations~\cite{copeland:2015}, and help quantify the odds of a large war over the next 100 years~\cite{cederman:weidmann:science:2017,braumoeller:2019}. 

% 2. 
Large armed conflicts tend to have disproportionate impact on political, economic, and social systems, and the larger the conflict, the broader those impacts tend to be worldwide. A single large war that kills 500,000 is clearly worse than 10 ``small'' wars that kill 5,000 each. This underscores the limitation of assessing the costs of war by counting incidence, without considering variations in their size. But beyond the direct impact, a large war that kills 500,000 is arguably also worse than a number of smaller wars, through its greater indirect consequences~\cite{coletal:2003}. Despite its importance, war size remains understudied both empirically and theoretically. In contrast, far more is known and theorized about how wars begin (onset)~\cite{colhoff1998,cgb:2013,mitchell:vasquez:2021} and how they end (termination)~\cite{cunetal:2009,reiter:2010}, which has informed efforts to prevent or reduce armed violence ~\cite{goertz:etal:2016,goldstein:2011}. Studies of war size tend to focus on characterizing the risk of the largest and most destructive conflicts~\cite{braumoeller:2019,clauset:2018,copeland:2001,levy:1983,richardson:1948} (including so-called ``major power'' wars). However, despite these advances, claims of lasting trends toward fewer or smaller conflicts remain controversial, particularly in the postwar period that began after the Second World War~\cite{goldstein:2011,roth:2018,braumoeller:2019}, and relatively little attention has been paid to the question of how wars become large in the first place.

% 3. 
Two key difficulties have limited progress on understanding how wars become large. One arises from the uncertain relationship between the broad variation in types of wars and their sizes. Most work on wars draws a clear distinction between civil wars, defined as armed conflicts between states and non-state actors, and interstate wars, defined as conflict between two sovereign states. But does war size accumulate differently in civil and interstate wars? Does the motive or incompatibility of a civil war, e.g., whether an ethnic or separatist conflict or effort to displace the central government, govern how large it may become? Empirically, civil wars and interstate wars differ in their typical size, frequency, and duration (Fig.~1). Since 1946, the widely used Correlates of War data~\cite{way:sar:2010} counts $38$ interstate wars worldwide, with an average size of $93,\!334$ battle deaths, with an average duration of $2.1$ years. In contrast, over the same period, it counts $5$ times as many civil wars ($191$ conflicts), which are $4.5$ times smaller ($20,\!858$ battle deaths on average) but $1.7$ times longer in duration ($3.5$ years on average). And, although interstate conflicts are far more likely than civil wars to end in a negotiated peace~\cite{pillar:1983,hegre:2004,fearon:2004}, it remains unclear how that tendency impacts their sizes. As a result, these differences between civil wars and interstate wars make it difficult to assess whether there are commonalities in how conflicts become large.

%% FIGURE 1 -- 
\begin{figure*}[t!]
\begin{center}
%--------------- FIGURE 1 ---------------
\includegraphics[scale=0.405]{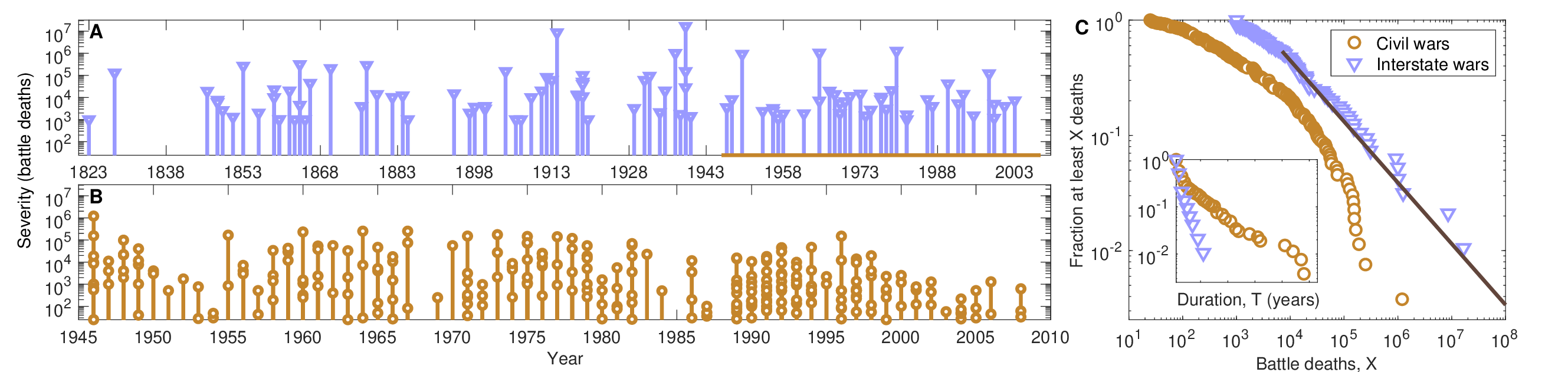}
\end{center}
\vspace{-3mm} \caption{\textbf{Conflict severity and duration.} Conflict severities (battle deaths) by year of onset for (A)~the 95 interstate wars 1823--2003 in the Correlates of War interstate war data~\cite{way:sar:2010} and (B)~264 civil wars and internationalized civil wars 1946--2008 in the PRIO Battle Deaths data~\cite{lacina:gleditsch:2005} (omitting conflict ongoing in 2008). The solid horizontal line in the upper panel highlights the range of time covered in the lower panel. (C)~Severity distributions for the same conflicts, along with the maximum likelihood power-law model for the largest-severity interstate wars (solid line, \mbox{$\hat{\alpha}=1.53\pm0.07$} for $x\geq7061$~\cite{clauset:2018}). (inset)~Distributions of conflict durations in years, showing civil wars' much longer durations despite their statistically lower severity. However, conflict severity only weakly correlates with conflict duration (all conflicts:\ $r^{2}=0.15$). }
\label{fig:1}
\end{figure*}
% ---

% 4. 
A second difficulty stems from a broad lack of disaggregated or within-conflict data on variations in fighting intensity~\cite{cederman:gleditsch:2009}. War size is commonly measured as the cumulative number of battle deaths over the duration of the conflict~\cite{richardson:1948,way:sar:2010,braumoeller:2019}. This focus on military deaths will undercount a war's total impact, including civilian deaths and indirect effects, which can extend far beyond the end of a military conflict~\cite{ghr:2003}, but has the practical advantage that military deaths tend to be more reliably counted. Still, a focus on aggregate severity in cumulative battle deaths tends to obscure differences in fighting intensity over time and in conflict duration, because longer wars have more time to accumulate casualties.

% 5. 
Here, we develop and test a generic model that connects variations over time in fighting intensity---a conflict's ``escalation dynamics''---with a conflict's duration to explain how wars grow to their final sizes, and we determine whether escalation dynamics can explain the observed variation in the sizes of modern civil and interstate wars. Our analysis relies heavily on the PRIO Battle Deaths data set~\cite{lacina:gleditsch:2005}, which provides comprehensive, disaggregated estimates of annual battle deaths in civil and interstate wars spanning 1946--2008. These data indicate that high-variance escalation dynamics are a generic feature of armed conflict, and the empirical risk of a conflict becoming 10 or even 100 times more deadly in the next year closely equals the likelihood of concomitant deescalation, except in the case of large civil wars, where we find a slight systematic tendency toward deescalation.

% 6. 
Combining escalation dynamics with empirical models of war onset and duration yields a non-parametric model that we use to determine whether escalation dynamics can explain the observed variation in civil and interstate war sizes. We find that both the sizes of civil wars in the postwar period and the scaling relation between size and duration can be fully reproduced by escalation dynamics, and moreover that civil war sizes cannot be explained without them. Under simple variations in the escalation model, we find that increasing the magnitude of fighting intensity can also closely reproduce the distribution of interstate war sizes over the past 200 years, and again, interstate war sizes cannot be explained without escalation dynamics. Finally, we explore the model's predictions in two forecasting tasks:\ estimating the size of potential future civil wars in four large nations, and estimating the final size of several civil wars that were ongoing in 2008. The results demonstrate that escalation dynamics drive enormous uncertainty in war size, even for conflicts that tend to begin small, and obtaining narrower estimates will require accounting better for how conflict-specific factors shape escalation dynamics.

% ------------------------------------------------------------------------------
\section*{Materials and Methods}

\noindent \textbf{War severity and duration}

\noindent 
% 7. 
Our analysis focuses narrowly on how conflict size or severity accumulates over a conflict's duration. We do not consider the declared reasons for conflict or the context in which it occurs, including the manner or locus of fighting, geography, military capacity or strategy, or relationships with other conflicts past or future. Nor do we consider how these additional dimensions of a conflict may themselves evolve over time. This simple focus on severity allows us to develop a generic model of within-conflict escalation dynamics that can later be adapted to account for conflict-specific factors.

% 8. 
Interstate war severity is well known to exhibit a strongly right-skewed pattern~\cite{clauset:2018}, such that the largest interstate wars are many orders of magnitude more severe and less frequent than a ``typical war'' (Fig.~1C). For instance, the Correlates of War data on wars since 1823 records 16,634,907 battle deaths for the Second World War, while the median size of an interstate war is only 7900 battle deaths~\cite{way:sar:2010}. Since L.~F.~Richardson's seminal work on conflict sizes in the mid-20th century, this broad variance in war sizes has often been described by a stationary power-law distribution~\cite{richardson:1948,clauset:2018}---called ``Richardson's Law''---which implies a small but enduring risk of large conflicts over the next century~\cite{clauset:2018}.

% 9. 
Civil war severity also exhibits a highly right-skewed pattern (Fig.~1C), albeit one that is shifted ``down'' in overall magnitude and with fewer very large civil wars compared to the interstate war pattern. For instance, the PRIO Battle Deaths data records the Chinese civil war as causing more than 1,000,000 battle deaths (from 1946 and onward), but the median for civil and internationalized civil wars in the postwar period is only 684 deaths~\cite{lacina:gleditsch:2005}. At the same time, civil wars can last far longer than interstate wars:\ although about half of interstate conflicts since 1823 and slightly more than half of civil wars since 1946 lasted no more than two years, the longest interstate war lasted 11 years while the longest civil war has lasted at least 63 years.

% 10. 
The substantial body of research on civil wars and how they may differ from interstate wars~\cite{mason:mitchell:2022} says little about war severity. At best, past studies have shown a poor correlation between factors associated with civil war onset and subsequent severity~\cite{lacina:2006}, and theoretical models of conflict dynamics like contest success functions or war-of-attrition models have not been applied to account for variation in conflict severity or to guide empirical research on severity~\cite{hirshleifer:1988}. Research on conflict duration and termination has noted clear differences between interstate and civil wars. Some have suggested that problems of credible commitment are more intractable in conflicts between states and non-state actors, leading to longer civil conflicts than interstate wars, which tend to be more amenable to negotiated settlements once the likely outcomes and battlefield performance are revealed after conflict onset~\cite{hegre:2004,fearon:2004,pillar:1983}. However, these models say little about how or when escalation might occur within a conflict, and, if anything, suggest that deescalation should be the norm due to the information gained from engagement.

% 11. 
Empirically, war severity and duration are relatively weakly correlated (see Appendix~A, Fig.~S1), implying that longer wars are not necessarily more severe. For example, as of 2008 the civil war between the Ethiopian government and the Oromi Liberation Front had been ongoing for 25 years, but with only 650 estimated total deaths, while the Chinese civil war's saw an extreme death toll over only 4 years (1946--1949). The lack of a clear theoretical explanation for war severities and the low correlation between war sizes and durations highlights the key question:\ how do large conflicts become so large? There are only three possible ways a conflict could become large:\ it begins with a large severity, it lasts long enough to accumulate many casualties, or the fighting intensity escalates.
\\

%% FIGURE 2 -- 
\begin{figure*}[t!]
\begin{center}
\begin{tabular}{ccc}
\includegraphics[scale=0.30]{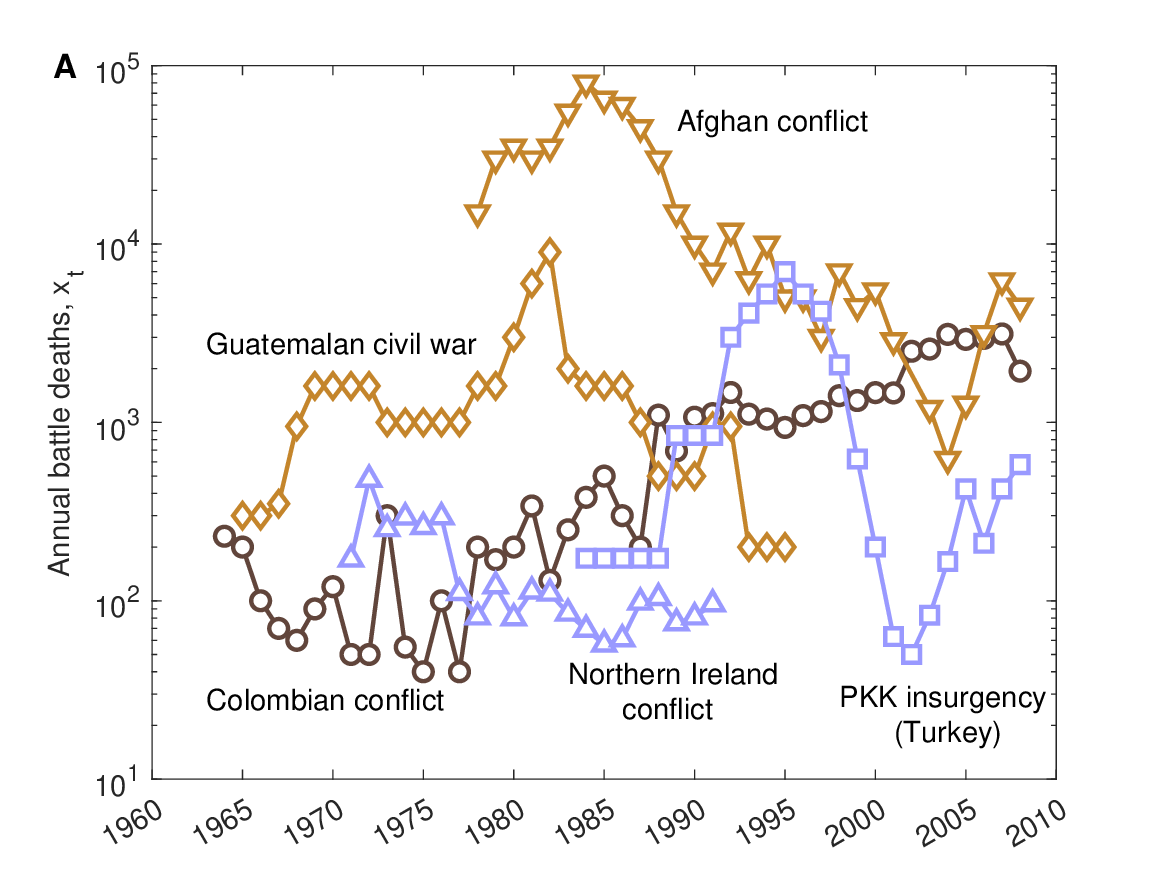} & 
\includegraphics[scale=0.30]{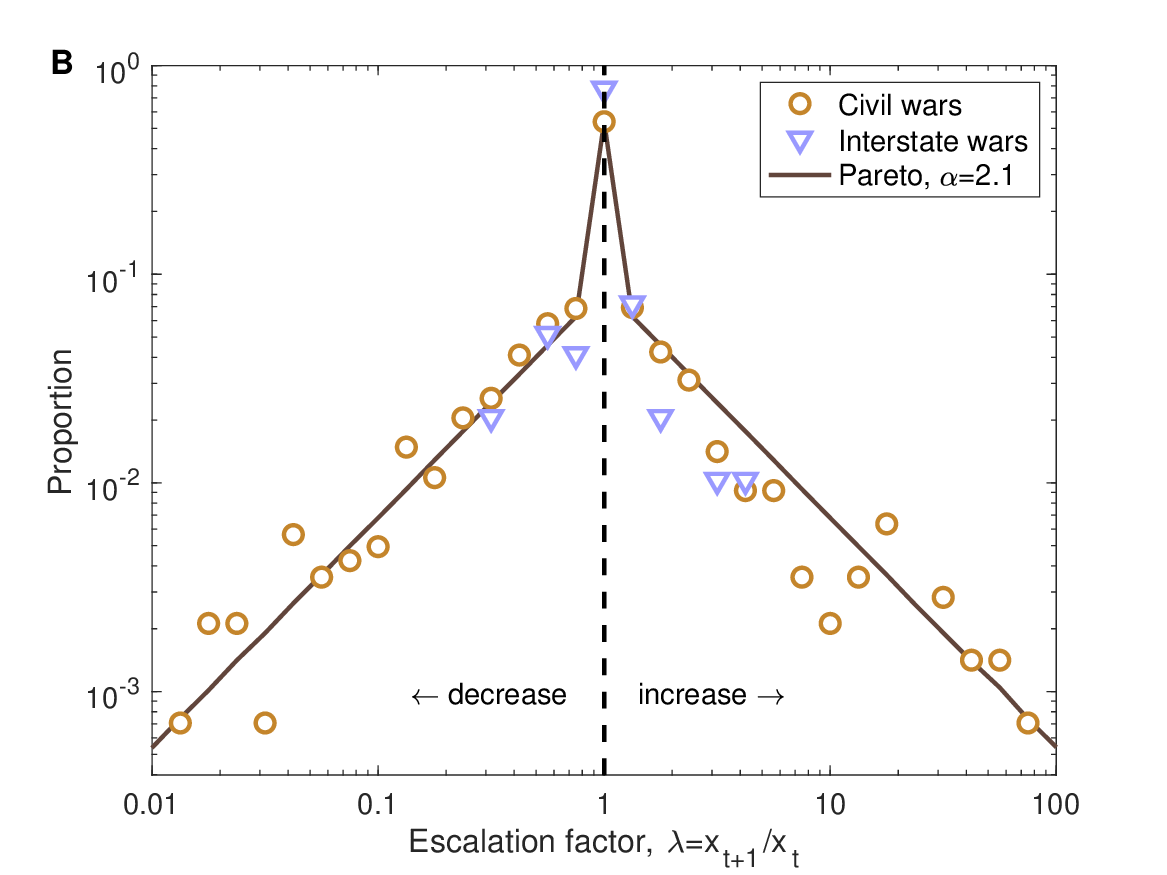} & 
\includegraphics[scale=0.30]{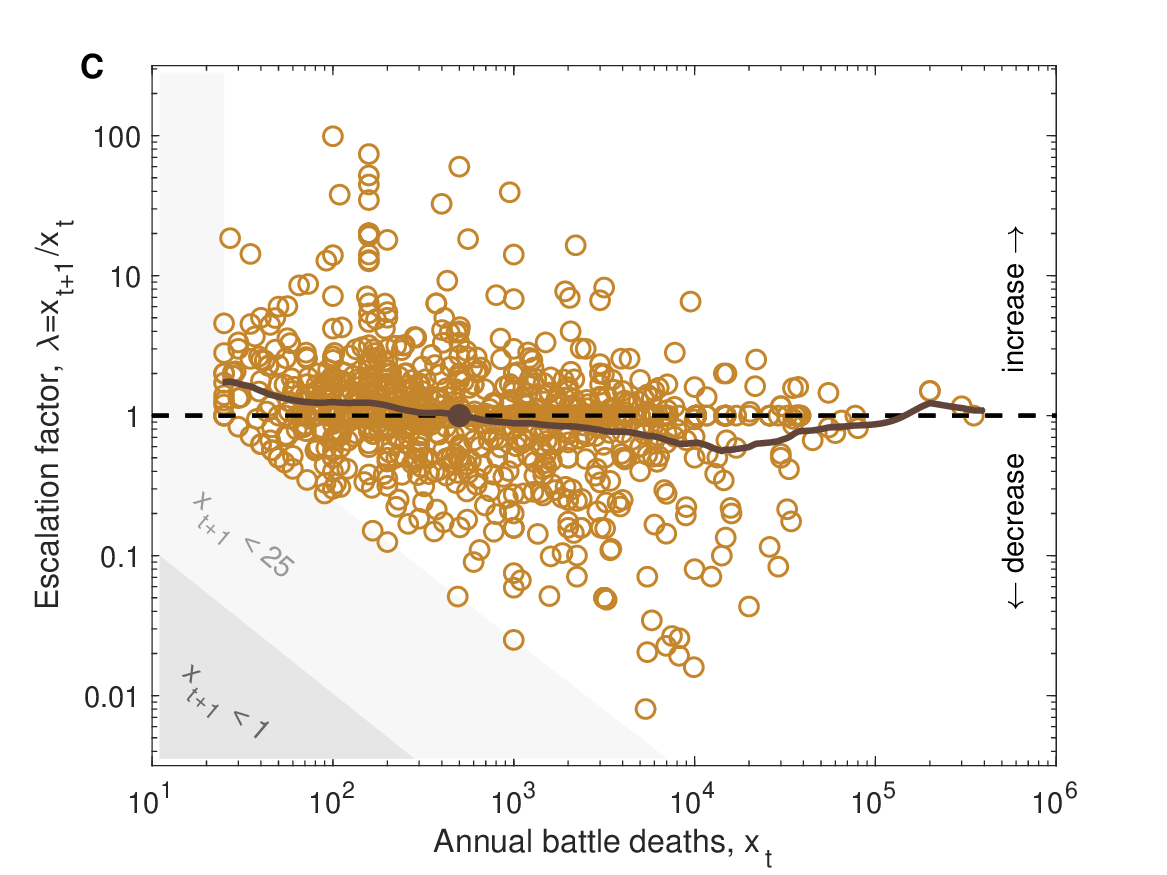} 
\end{tabular}
\end{center}
\caption{\textbf{Civil war severity dynamics.} (A)~Severity time series $x_{t}$ for five selected civil war conflicts, showing highly variable periods of escalation and deescalation in annual battle deaths $x_{t}$ over time. (B)~Distribution of escalation factors $\Pr(\lambda)$, where $\lambda_{t} = x_{t+1}/x_{t}$, which defines the likelihood of a conflict increasing, or decreasing, it annual severity by a factor $\lambda$ in the next year, for both civil wars (circles; $n_{\lambda}=1415$, from 188 conflicts with $T>1$) and interstate wars (triangles; $n_{\lambda}=97$, from 18 conflicts with $T>1$) since 1946, along with a double Pareto distribution with parameter $\hat{\alpha}=2.0\pm0.1$. (C)~Joint distribution $\Pr(\lambda\,|\,x_{t})$ of civil war escalation factors $\lambda_{t}$ vs.\ current annual conflict severity $x_{t}$, showing a systematic correlation between escalation and severity, in which conflicts below (above) $x_{t}=500$ tend to escalate (deescalate) in the next year. Shaded regions indicate `'forbidden' factors due to minimum measurable severity.}
\label{fig:2}
\end{figure*}
% ---

\noindent \textbf{Data and statistical models}

\noindent 
% 12. 
The dynamics of fighting intensity over the course of an armed conflict can be represented as a time series of annual severities $x_{1},x_{2},\dots,x_{T}$, where $x_{t}$ is the severity (battle deaths) in year $t$, $T$ is the conflict's total duration in years, and $X=\sum_{t} x_{t}$ is the conflict's cumulative size (Fig.~2A). Equivalently, the same conflict can be represented by a combination of $x_{1}$, the severity of the first year of fighting, and a sequence of ``escalation factors'' $\lambda_{t}=x_{t+1}/x_{t}$ that denote the proportional increases (escalation, $\lambda>1$) or decreases (deescalation, $\lambda<1$) in annual severity each year. These escalation factors measure the changes in levels of fighting, rather than measure the absolute levels themselves, thereby emphasizing the conflict's dynamics rather than its steady state.

By treating a conflict time series as generated by a first-order stochastic process, each conflict year's severity is a product of the current year's severity and the degree of escalation:\ $x_{t+1}=\lambda\,x_{t}$. Consecutive years of $\lambda>1$ would indicate a period of repeated escalation, driving a conflict toward larger sizes, while consecutive years of $\lambda<1$ will have the reverse effect (deescalation). Moreover, the distribution of escalation factors across conflicts would reveal any systematic tendency toward escalation or deescalation, and would quantify the likelihood of different levels of escalation and deescalation.

% 13. 
We extract severity time series for all conflicts worldwide from 1946--2008 contained in the PRIO Battle Deaths data (PBD) v3.1~\cite{lacina:gleditsch:2005}. 
We consider all years of armed violence for (i)~civil wars and internationalized civil wars, of which $n=299$ meet our inclusion criteria (see Appendix~B), resulting in $1714$ conflict-years, and separately (ii)~interstate wars and extra-state wars, of which $n=22$ meet our inclusion criteria, resulting in $119$ conflict-years. We omit conflicts coded as extra-systemic (conflicts in colonies, outside independent states), and we split each included conflict into time series of continuous armed fighting with no more than 1 year of below-threshold severity. The resulting data set is disaggregated below the level of entire conflicts, but remains aggregated at the level of entire years. Hence, annual severity $x_{t}$ may not be an accurate measure of typical conflict severity at the level of weeks or days, across space, or within individual battles.

When analyzing the total severity of interstate wars, we use to the 95 such conflicts over the 1823--2003 period contained in the Correlates of War (CoW) v4 interstate conflict data set~\cite{way:sar:2010,smallsing:1982}. This data set provides comprehensive coverage in this period, is widely understood and well-scrutinized, and is the most common reference set for claims about the sizes of interstate wars. To consider differences in the size of countries we use annual country population estimates from the United Nations Population Division~\cite{unpopdiv}, which are based on national population censuses as well as estimates and projections onward to 2100.

% ------------------------------------------------------------------------------
\section*{Results}

\noindent \textbf{Escalation dynamics in armed conflicts}

% 14. 
\noindent
Substantial conflict escalation is likely to require a ratcheting-up in conflict-waging effort. Repeated escalation over multiple years should thus pose significant political and logistical difficulties from repeatedly building new capacity or continually reallocating resources away from civilian needs. Distinct difficulties are likely to apply to repeated conflict deescalation. Thus, substantial escalation or deescalation in fighting should be punctuated or occasional phenomena rather than continuous, and empirical escalation factors should tend to cluster around $\lambda=1$ (no-change).

% 15. 
For instance, in the Afghanistan conflict (Fig.~2A), annual severity was high and relatively steady prior to 1990, when the government supported by the Soviet Union was fighting rebels heavily armed and trained by the United States, killing around 30,000 per year. From 1990, after Soviet forces withdrew, annual severity fell to a substantially lower level, around 5000 per year, reflecting the lessened direct foreign state support to both the government and the non-state actors. In contrast, conflict severity in the Guatemalan civil war (1963-1995) was relatively stable over most of its duration, with around 2000 deaths per year, except for a brief period in the early 1980s where fighting spiked to 10,000 deaths per year before returning to its previous level.

% 16. 
For civil wars with durations $T>1$ year, empirical escalation factors across all conflict years exhibit a highly symmetric distribution that is peaked at $\lambda=1$ (Fig.~2B) and we find no evidence of a systematic tendency to escalate or deescalate. Instead, we find that (i)~the most commonly observed change in severity within a conflict (44\% of all escalation factors) is no-change, i.e., fighting holds steady, and (ii)~the likelihood that a conflict's annual severity may increase to be 10 or even 100 times larger in the next year closely follows the likelihood of a concomitant decrease in severity. This unconditional distribution ignores any correlations among escalation factors or correlations with conflict covariates or the conflict's current severity, a point we return to below. 
We note that the abundance of $\lambda=1$ escalation factors is likely an artifact of the PBD's construction:\ for many multi-year periods within conflicts, or for some entire conflicts, the annual severities simply equal the period's total divided by its length. However, there is no reason to expect that correcting this artifact would alter the shape of the distribution's tails, i.e., the tendency to escalate or deescalate.

% 17. 
We characterize the shape of the distribution of escalation factors using a piecewise model, in which with probability $q$, there is no change in severity this year ($\lambda=1$), and otherwise $\lambda$ is drawn from a double Pareto distribution, which has symmetric power-law tails above and below the modal value. Using standard statistical techniques~\cite{clausetetal:2009}, we find that the maximum likelihood power-law tail parameter is $\hat{\alpha}=2.1\pm0.1$, indicating an extremely high variance distribution~\cite{newman:2005}. Furthermore, we cannot reject the Pareto distribution as a data generating process for the tails of the escalation factor distribution ($p_{\rm KS}=0.31\pm0.03)$. That is, the observed escalation factors are, as a group, statistically indistinguishable from an iid draw from the fitted double Pareto distribution. 

% 18. 
Interstate wars do not permit an independent characterization of their escalation factor distribution because of their smaller number in the PBD period (1945-2008) and shorter durations. However, we find that interstate war escalation factors (excluding $\lambda=1$) are statistically indistinguishable from the empirical distribution of civil war factors (2-tailed KS test, $p=0.16$), and are a plausible iid draw from the estimated model of civil war factors ($p_{\rm KS}=0.61\pm0.03$). That is, we find no evidence that escalation dynamics differ significantly between civil and interstate wars, even as the cumulative sizes and durations of these conflicts differ substantially (Fig.~1C), suggesting that high-variance escalation dynamics are a generic feature of armed conflict.

% 19. 
Within civil wars, the escalation factors correlate slightly with annual severity, such that ``hot'' conflicts tend to deescalate in the next year, while ``cold'' conflicts tend to escalate (Fig.~2C). The effect at the lower end is attributable to left-censoring, because small conflicts can only become so much smaller before their severity falls below a measurable threshold (in the PBD, $x_{\textrm{min}}=25$). Hence, conditioned on conflict continuing for another year at measurable severities, an otherwise symmetric distribution of conflict escalation factors (Fig.~2A) would be truncated on the left side, shifting the average escalation factor to be $\langle \lambda \rangle >1$ (escalation). Although there is no such constraint on the upper end, we nevertheless observe a symmetric tendency toward deescalation among large civil wars, i.e., $\langle \lambda \rangle <1$ (but no such tendency among large interstate wars Fig.~S2). The cross-over point between these two regimes occurs at about $x_{t}=500$. Hence, civil wars will tend to regress, over time, toward this ``warm'' value of severity, although the broad variance of escalation factors will tend to obscure that pattern in any particular conflict. An important direction of future work is to understand the causes of this systematic tendency for large civil wars to deescalate, which could be related to fundamental constraints arising from population, military capabilities, recruitment, bargaining, war fatigue, or even international pressure.
\\

\noindent \textbf{The severity of civil wars}

% 20. 
\noindent
Escalation dynamics highlight how fighting intensity often changes over the course of a conflict, and our results indicate that escalation is a common pattern in armed conflicts. We now develop a generic model of severity dynamics, based on escalation factors, and determine whether escalation dynamics can explain the cumulative sizes of civil wars (Fig.~1C).

%% FIGURE 3
\begin{figure}[t!]
\begin{center}
\includegraphics[scale=0.45]{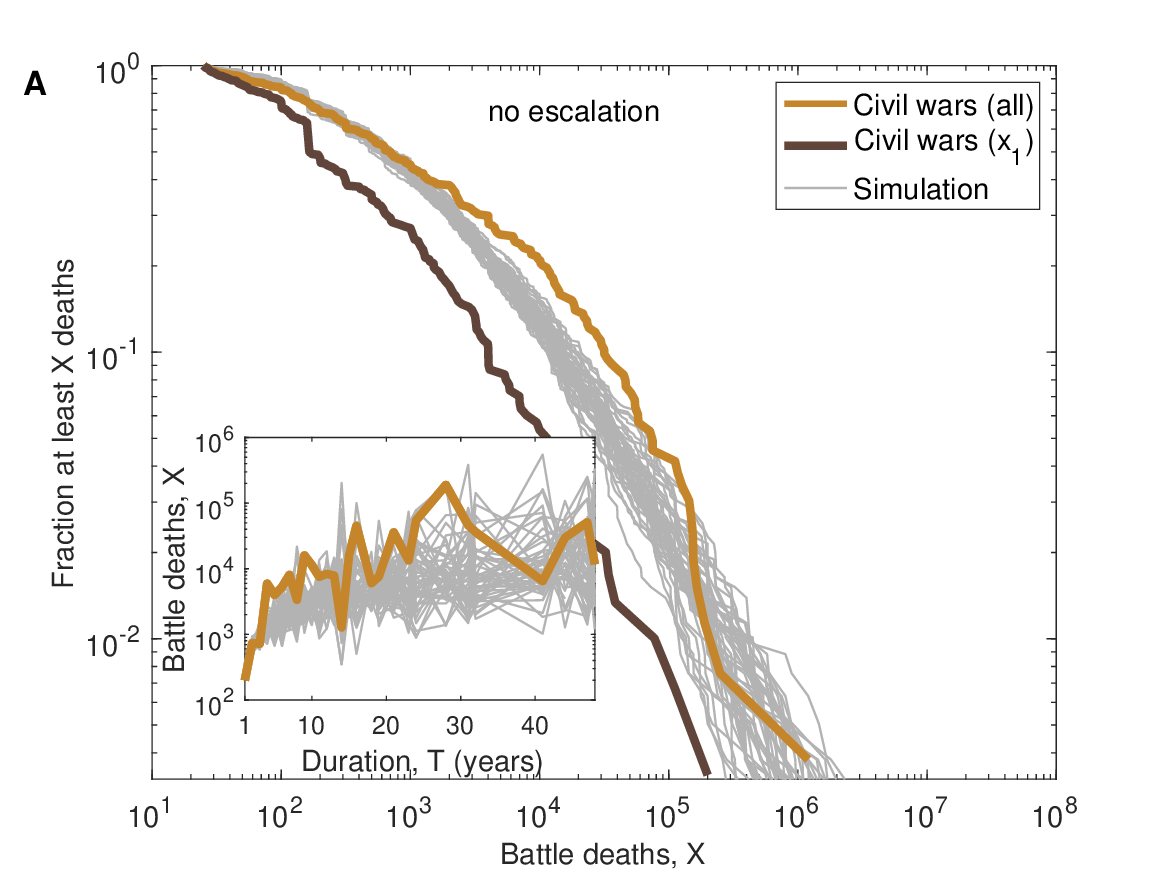} \\
\includegraphics[scale=0.45]{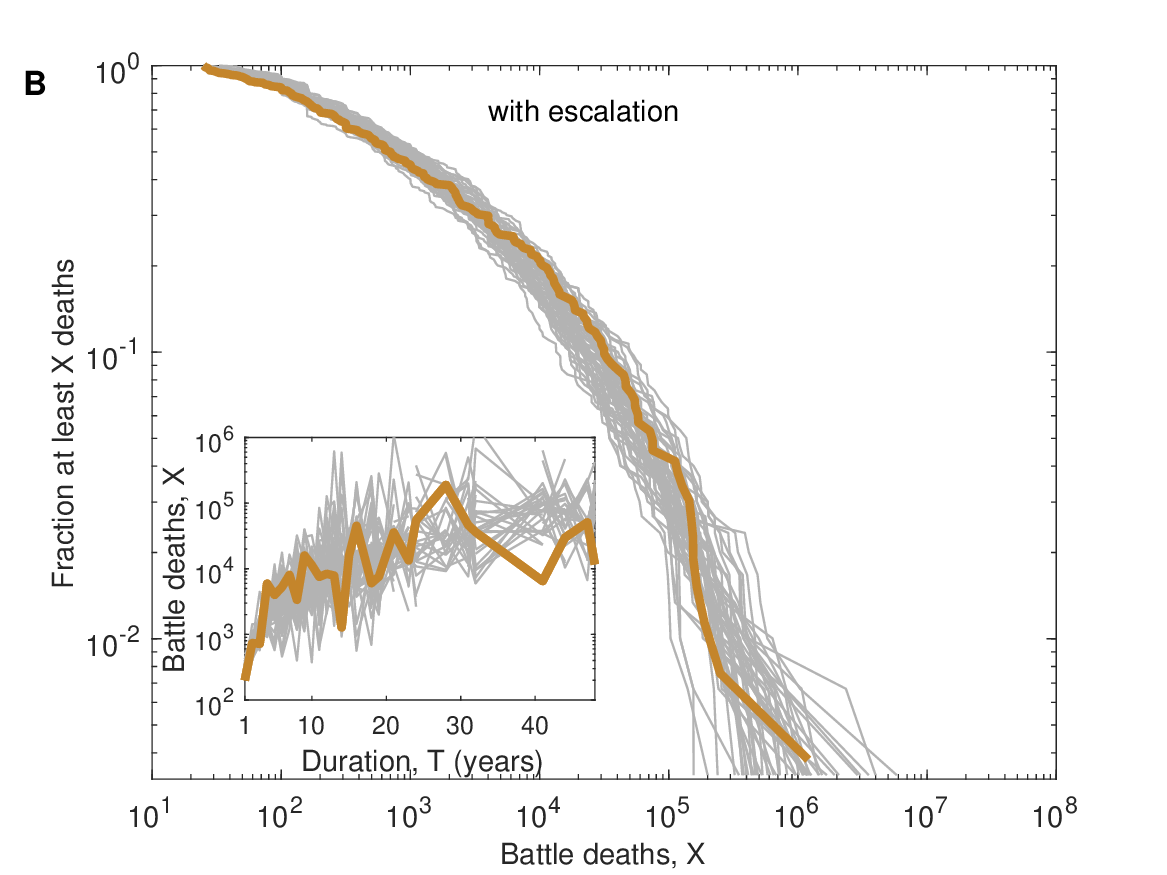} 
\end{center}
\caption{\textbf{Simulating civil war severities}. (A)~Total severities (battle deaths) for civil wars 1946--2008, along with conflict initial severities $x_{1}$ and 50 simulations of total severity, with effects from initial severity and duration but without escalation. (B)~Total severities for cvil wars, along with simulated severities from the non-parametric escalation model (see text), showing close agreement. Insets:\ mean conflict severity $\langle X \rangle$ as a function of conflict duration $T$ (years) for civil wars and the corresponding simulations, showing closer agreement under the escalation model.}
\label{fig:3}
\end{figure}
% ---

%% FIGURE 4 -- 
\begin{figure*}[t!]
\begin{center}
\begin{tabular}{ccc}
\includegraphics[scale=0.30]{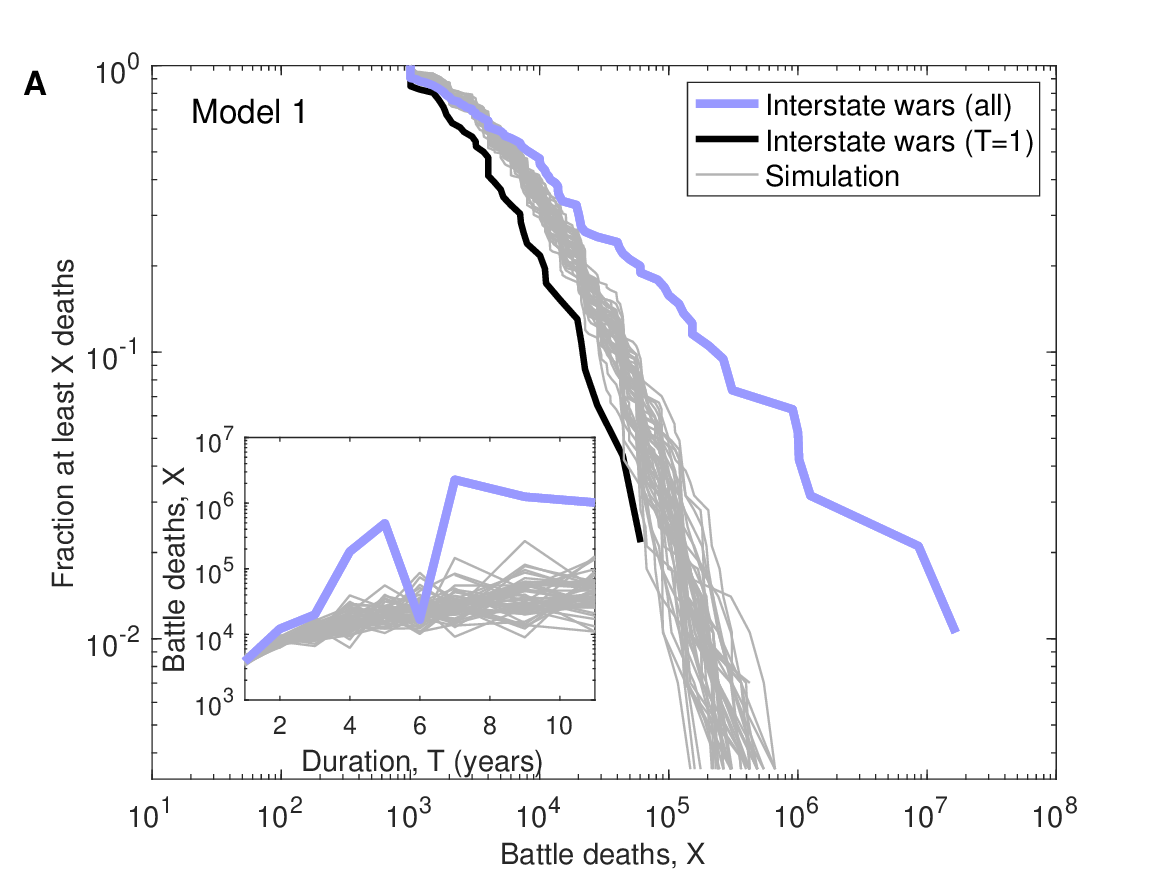} &
\includegraphics[scale=0.30]{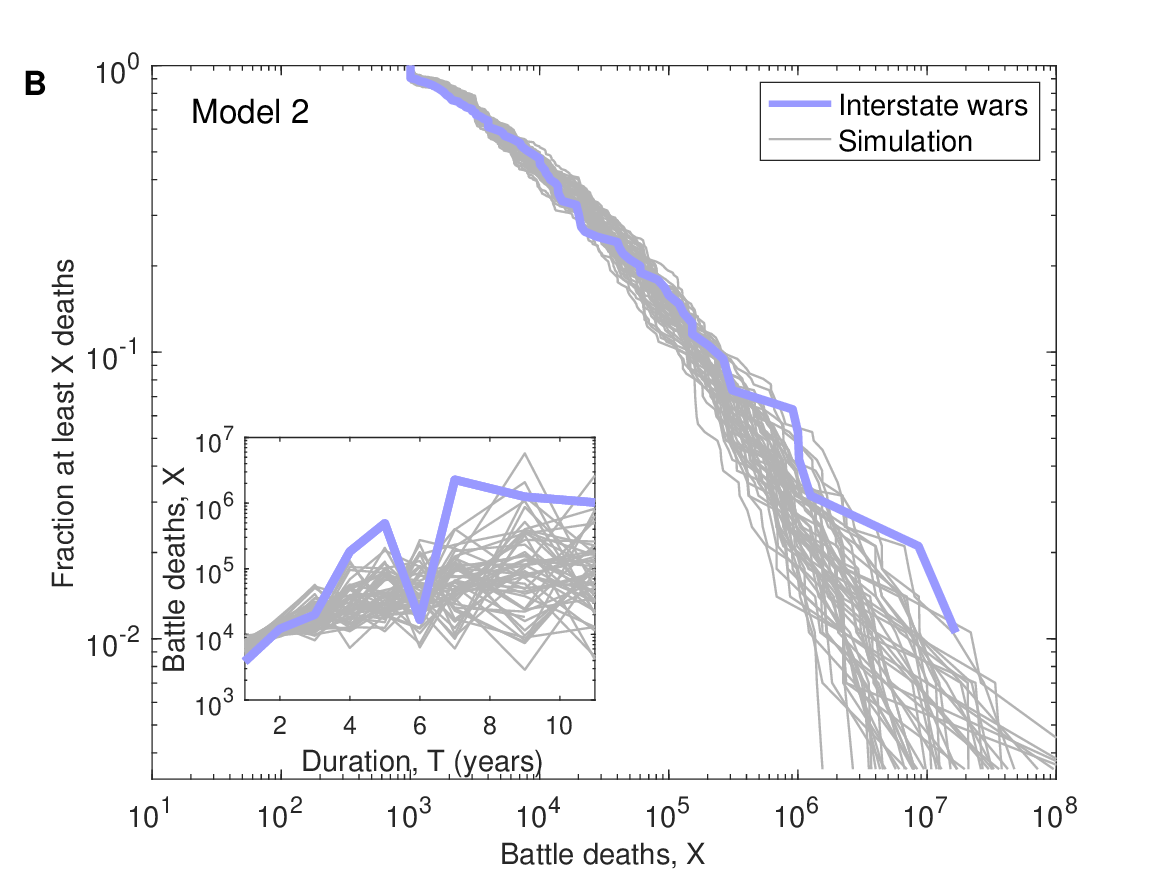} &
\includegraphics[scale=0.30]{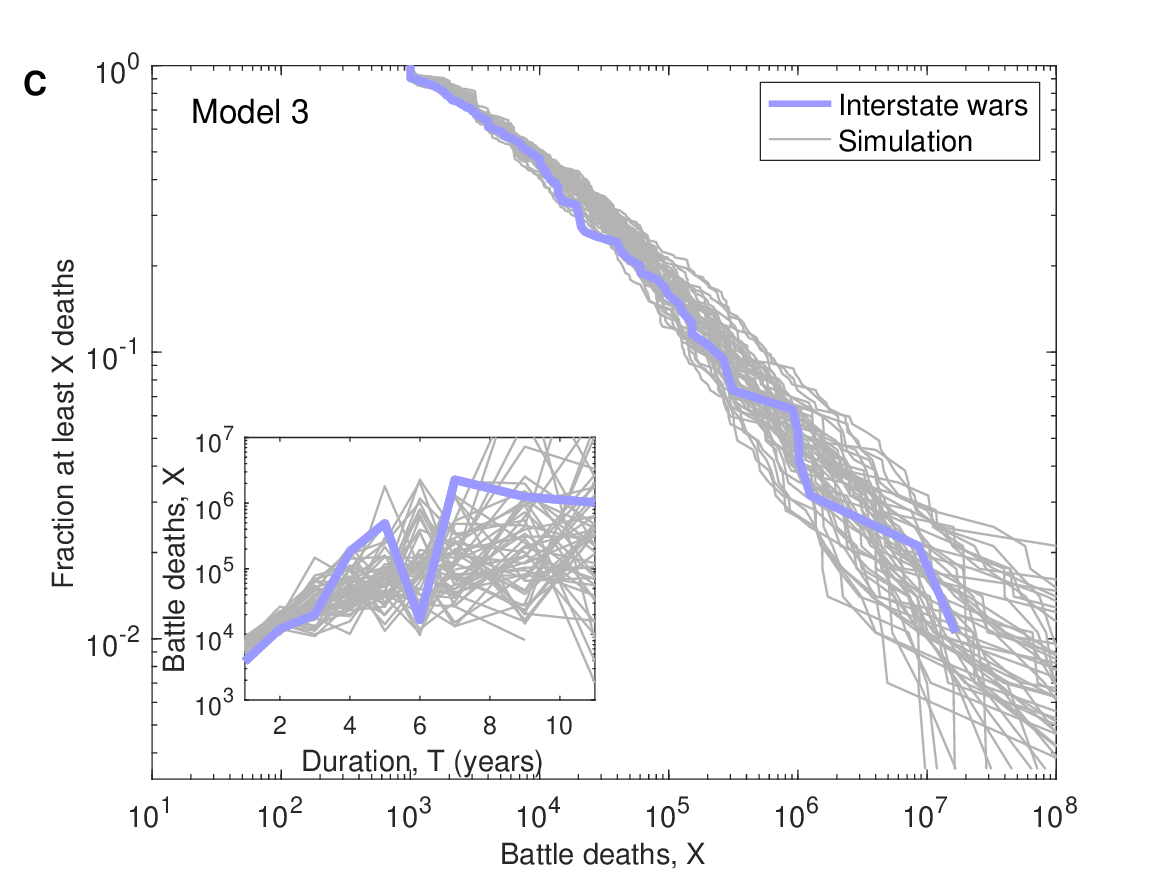} 
\end{tabular}
\end{center}
\caption{\textbf{Simulating interstate war severities}. (A) Total severities (battle deaths) for interstate wars 1823--2003, along with severities for the subset of $T\!=\!1$ year duration conflicts and 50 simulations of total severities derived from Model~1, with effects from initial severity and duration but without escalation. (B,C) Total severities for interstate wars, along with simulated severities under Models~2 and~3 (see text), indicating that a shift in scale alone (Model~2) produces moderate agreement, while a shift in escalation (Model~3) produces better agreement. Insets:\ mean conflict severity $\langle X \rangle$ as a function of conflict duration $T$ (years) for interstate wars and the corresponding Model simulations, showing closest agreement under Model~3. }
\label{fig:4}
\end{figure*}
% ---

% 21a. 
The escalation model generates a conflict time series $x_{1},x_{2},\dots,x_{T}$ in three parts:\ conflict duration, initial severity, and escalation dynamics. First, we draw the conflict duration $T$ uniformly at random from the empirical distribution of civil war durations $\Pr(T)$ (Fig.~1C inset). Second, we draw the severity in the first year of fighting $x_{1}$ uniformly at random from the empirical distribution of initial severities $\Pr(x_{1})$. Third, for each year $1 \leq t < T$, we draw an escalation factor $\lambda_{t}$ uniformly at random from the joint distribution of civil war escalation factors and current severity $\Pr(\lambda \,|\, x_{t})$ (see Appendix~C) and record $x_{t+1} = \lambda_{t}\,x_{t}$. The simulated conflict's total severity $X$ is the sum of these annual severities. The escalation model is fully non-parametric, with no fitted parameters, and instead combines the empirical distributions for duration, initial severity, and escalation factors into a simple random walk model of conflict annual severity.

% 21b. 
We contrast this escalation model with a ``no escalation'' null model that includes effects for duration and initial severity, but omits any effect from escalation. In this model, we draw the conflict duration~$T$ and initial severity~$x_{1}$ as in the escalation model, and then multiply the initial severity by the duration~$T$ to obtain the simulated total severity~$X$ (this process is equivalent to setting $\lambda_{t}=1$ for all~$t$). 

% 22a. 
The no-escalation model poorly reproduces the observed variation in civil war size over the period 1946--2008 (Fig.~3A), but does so in an interesting way. Initial sizes and durations alone do produce a sufficient number of both very small and very large civil wars, but they produce too few conflicts of intermediate size (those between 1000 and 100,000 battle deaths). Similarly, this model tends to produce conflicts that are systematically smaller in size $X$ for their given duration $T$ than the empirical data (Fig.~3A inset). Although the frequency of the very largest conflicts under the model agrees with the historical record, this agreement is misleading:\ the model lacks the empirical tendency for large civil wars to deescalate (Fig.~2C), and hence the agreement in the upper tail of the distribution reflects an implicit overestimate of the intensity and duration of these largest conflicts. 

% 22b. 
In contrast, the escalation model closely matches the entire distribution of civil war sizes, reproducing both the overall shape of the distribution, the frequency of intermediate-sized conflicts, and the frequency of the very largest conflicts (Fig.~3B). The escalation model also reproduces the observed pattern in how conflict size $X$ tends to increase with conflict duration $T$ (Fig.~3B inset). Hence, escalation dynamics appear to be essential for explaining the sizes of civil wars.
\\

%% FIGURE 5 -- 
\begin{figure*}[t!]
\begin{center}
\begin{tabular}{ccc}
\includegraphics[scale=0.30]{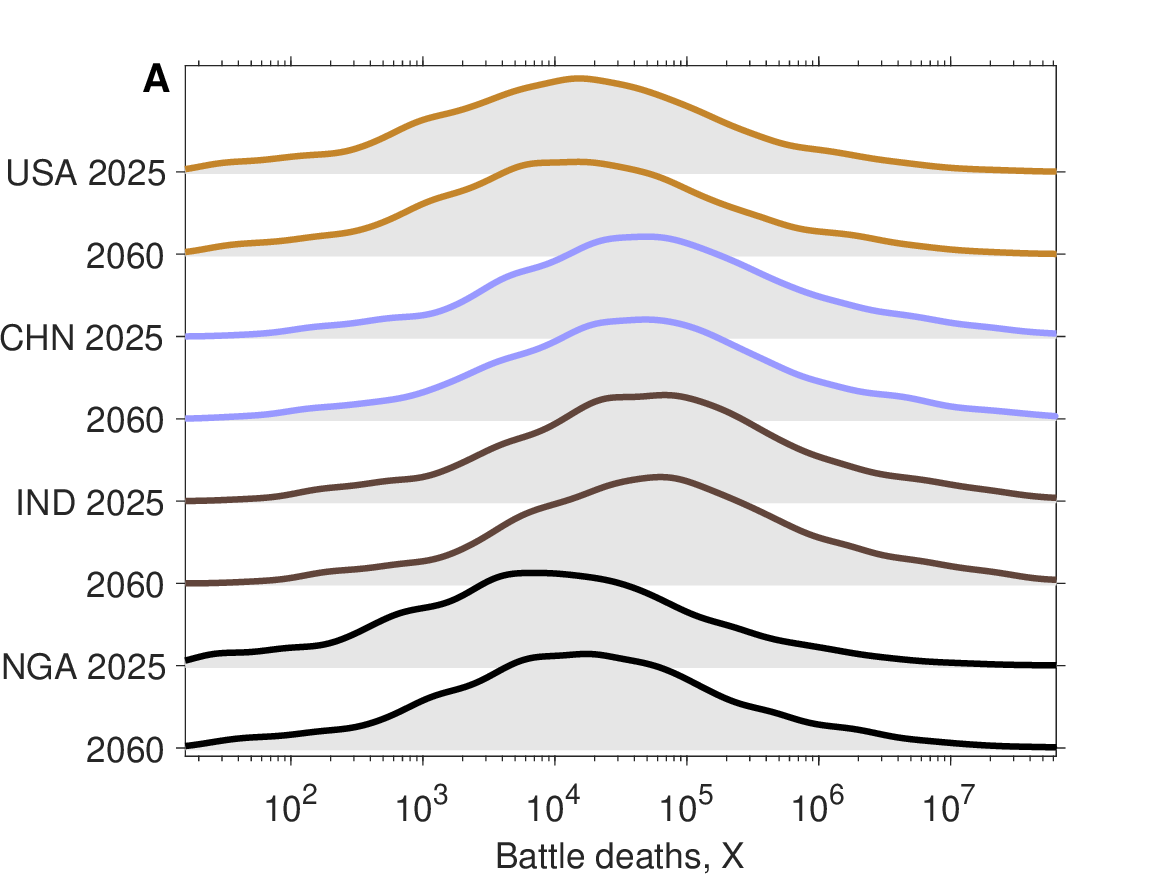} &
\includegraphics[scale=0.30]{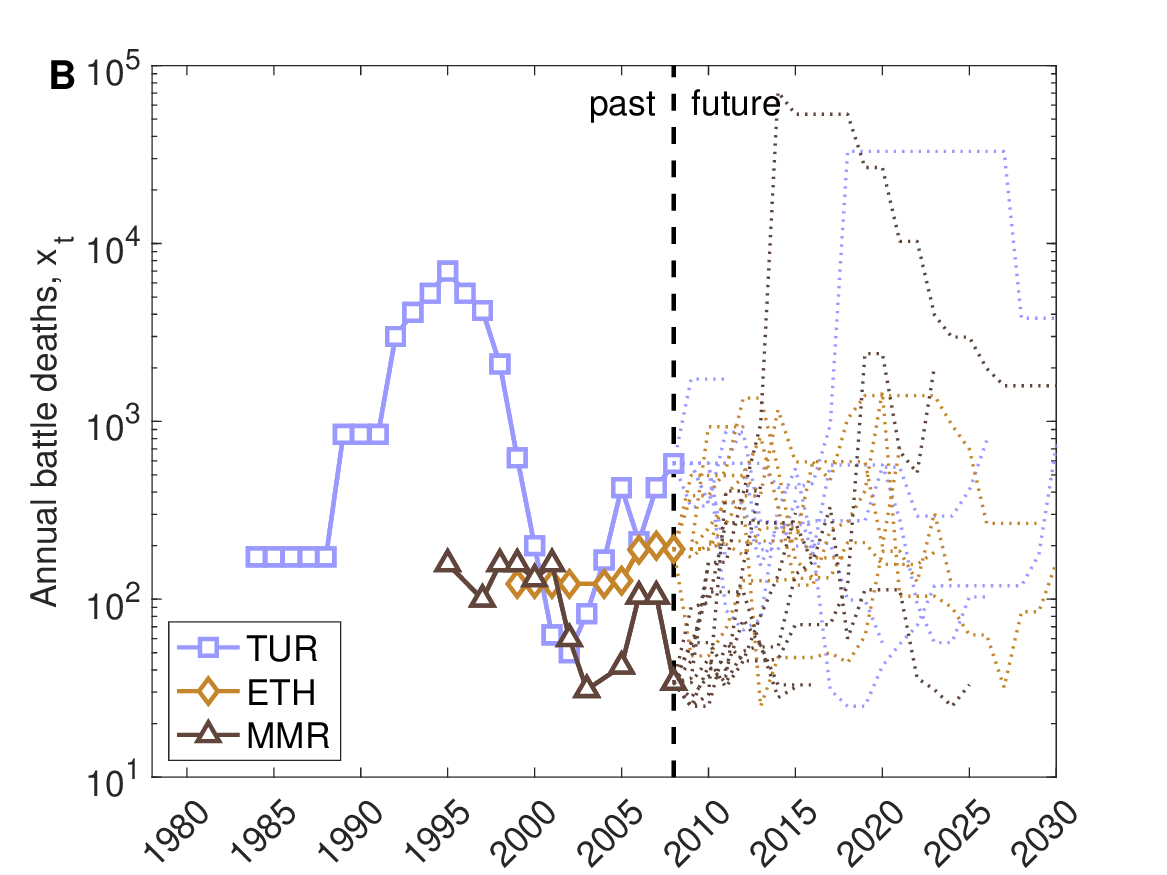} &
\includegraphics[scale=0.30]{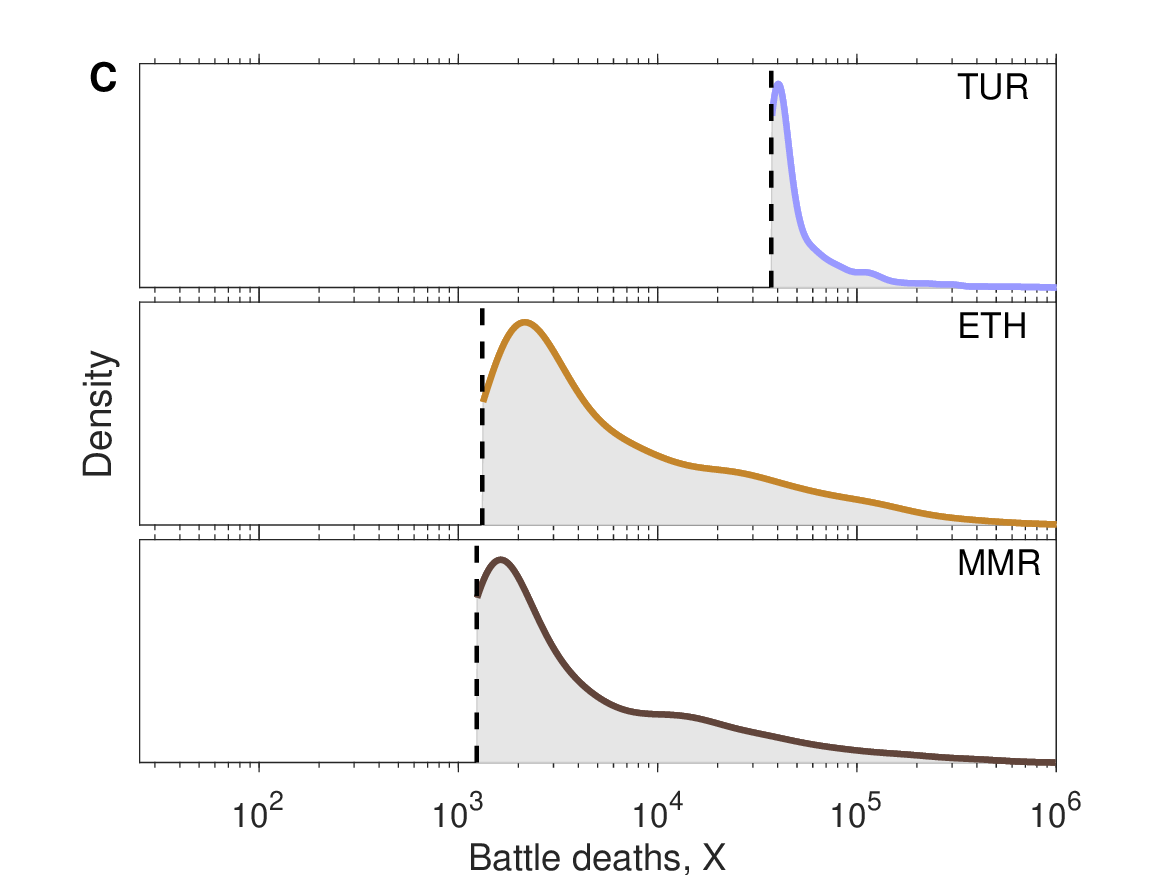}
\end{tabular}
\end{center}
\caption{\textbf{Forecasting civil war severity.} (A)~Distributions of simulated total severity $\Pr(X)$ (battle deaths) under the escalation model for hypothetical civil wars beginning in 2025 or 2060 in the United States of America (USA), China (CHN), India (IND), and Nigeria (NGA), in which the country's forecasted population in that year is used to generate the initial annual severity $x_{1}$ (see text). (B)~Severity time series $x_{t}$ for three civil wars, in Turkey, Ethiopia, and Myanmar, that were ongoing in 2008, showing both historical severities and 10 simulated future severity trajectories for each conflict, generated by the escalation model (see text). (C)~Distributions of simulated total severities for the same ongoing conflicts, with each conflict's cumulative severity as of 2008 indicated by a vertical dashed line. }
\label{fig:5}
\end{figure*}
% ---

\noindent \textbf{The severity of interstate wars}

% 23. 
\noindent
Testing the ability of escalation dynamics to explain the sizes of interstate wars requires a different approach, because we lack sufficient within-conflict information on interstate wars in the PBD to define an equivalent nonparametric model. Instead, we adapt the civil war model to interstate conflicts by defining and testing several variations to identify the sufficient conditions for generating large interstate wars. We evaluate these models using the severities of the 95 interstate conflicts in the Correlates of War data~\cite{smallsing:1982}.

% 24. 
Unless otherwise stated, all versions of the interstate war escalation model have two modifications relative to the civil war escalation model. First, because interstate wars tend to be substantially shorter in duration than civil wars, in the interstate escalation models, we instead draw a conflict's duration~$T$ from the empirical duration distribution for interstate conflicts. Second, we do not expect the tendency for large civil wars to deescalate to also appear in interstate conflicts, in which the belligerents are state actors with much greater capacity for mobilizing resources and hence are less subject to the self-limiting tendencies non-state actors experience in fighting civil wars. For instance, state-level belligerents can mobilize greater resources through taxation and conscription, and wars can expand to additional states, e.g., via alliances or geographic proximity~\cite{siverson:starr:1991,li:cranmer:2017}. Hence, for models with escalation, we draw escalation factors $\lambda_{t}$ from the unconditioned distribution of escalation factors $\Pr(\lambda)$, instead of the size-conditioned one.

% 25. 
In Model~1, we again include effects only for duration and initial severity, and omit the effects of escalation. Because we lack data on war severity in the initial year of interstate conflicts in the CoW data, we instead use the total severity of wars that lasted only 1~year in duration as a proxy. Hence, in Model~1, we draw the initial severity in this way and multiply it by the drawn duration. In Model~2, we select the initial severity by using the civil war distribution of initial severities $\Pr(x_{1})$ and then multiply the drawn initial size by fixed factor of 20 to capture the larger baseline size of interstate conflicts. Escalation factors are then drawn from the unconditioned distribution as described above. In Model~3, we scale up the initial severities in the same way as Model~2, and then draw two escalation factors for each increment of time in the simulation. This modification captures the idea that escalation dynamics in interstate conflicts unfold at a faster time scale than in civil wars.

% 26. 
Model~1, which omits escalation, fails to produce large interstate conflicts and fails to produce large enough conflicts for their durations (Fig.~4A), indicating that, like for civil wars, escalation dynamics are essential for producing large conflicts. In contrast, Model~2, which includes escalation, produces very heavy-tailed distributions in final conflict sizes. However, this model produces somewhat too few of the very largest conflicts, and conflicts tend to be slightly smaller than expected for their duration, relative to the empirical data (Fig.~4B). Model~3, however, closely matches both the observed sizes of interstate wars and largely captures the observed relation between conflict size and duration (Fig.~4C). (See Appendix~C for additional simulation results.)

% 27. 
The key component of Models~2 and~3 is the unconditioned distribution of escalation factors $\Pr(\lambda)$, which gives large conflicts an equal chance of further escalating or deescalating. In fact, Model~2 includes little other than this feature, and produces a distribution of conflict sizes that is surprisingly close to the empirical data, indicating that the asymmetric tendency of large civil wars to deescalate is the key difference between civil and interstate war sizes. The better agreement of Model~3 among the very largest conflicts suggests that more rapid escalation dynamics is sufficient mechanism for producing the largest conflicts. (We note that other variations of the model can also replicate this empirical pattern; see Appendix~C.)
\\

\noindent \textbf{Forecasting civil war severity}

% 28. 
\noindent If escalation dynamics accurately capture how the intensity of fighting can vary within an ongoing conflict, it may conceivably be used to make model-based forecasts of how large a current or potential future conflict may become. We consider two such forecasting tasks for civil war escalation. In the first, we consider hypothetical civil wars in four large states (the United States, China, India, and Nigeria) beginning in 2025 and in 2060, and we forecast the cumulative size of the hypothetical conflict. In the second, we consider three civil wars that were ongoing in 2008 (the last year of the PBD data), in Turkey, Ethiopia, and Myanmar, and forecast their eventual total duration and cumulative severity.

% 29. 
In each case, the primary difference among these forecasts is the escalation model's initial severity $x_{1}$, which sets the initial scale of the conflict; the severity of subsequent years are governed by the generic model of civil war escalation. For the hypothetical civil wars, we base the initial severity on the country's estimated population in the year of conflict onset, which we multiply by a random ``intensity'' factor $\gamma$ to produce the initial size of the new conflict. To make these initial severities reflect historical patterns, we calculate the empirical distribution of intensity factors from the PBD data using the recorded initial severities, which we match to UN\ estimates of state populations in that first year of conflict~\cite{unpopdiv}. For instance, the PKK insurgency in Turkey becomes a civil war in 1984 with 442 battle deaths relative to a population of about 47.6 million, implying an intensity factor of $442/47569000$. For the ongoing civil wars, we set the initial severity $x_{0}$ to be the severity in 2008. To select their future durations, we employ a simple coin-flipping model using a hazard-rate model estimated from the historical civil war duration distribution (see Appendix~D).

% 30. 
Escalation dynamics drive broad uncertainty in the severity forecasts of all four hypothetical conflicts (Fig.~5A). In the largest states (China and India), these forecasts have a modal value around 50,000 battle deaths, while in the smaller states (United States and Nigeria), the forecasts produce a moderately smaller modal value around 10,000 battle deaths. However, the distribution of sizes is very broad, with most of the density ranging from 1000 deaths to 1,000,000 deaths, a difference of 3 orders of magnitude. This broad variance illustrates the way in which escalation dynamics tends to amplify uncertainty in future severity, such that the size of a civil war can be broadly uncorrelated with the size of the state, i.e., a smaller nation can produce a larger civil war than a more populous nation, even as large nations have greater potential for large conflicts.

% 31. 
Similarly, the escalation model produces broadly distributed forecasts for ongoing conflicts, with some future severity time series exhibits dramatic escalation, while others exhibit far less variation (Fig.~5B). The resulting distributions of cumulative severity for these conflicts are broad, again highlighting the inherent uncertainty caused by conflict escalation dynamics (Fig.~5C).

% ------------------------------------------------------------------------------
\section*{Discussion}

% 29. 
\noindent Although large wars are dangerous precisely because of their disproportionate social, economic, and political costs, war size remains understudied both theoretically and empirically. As a result, war size often plays an unstated role as a measure of cost within existing theories of war deterrence, onset, and termination. For instance, the larger (more costly) a war may be, e.g., because of alliances or armaments, the more extreme the circumstances must likely become for rational actors to consider it, and, the larger (more costly) a war becomes once begun, the more likely it may be to end, e.g., for the purely economic reason that fighting is expensive. When war size is considered, the focus is typically on cumulative battle deaths, the total after fighting has ended, without considering how the conflict achieved its cumulative size, or how long it took to get there. Without a clear understanding of how large wars become large in the first place, debates about the likelihood of large and destructive conflicts remain incomplete and the risks uncertain.

% 30. 
There are three ways an armed conflict can become large:\ it can be large at the conflict's onset; it can last long enough to accumulate a large size; or, it can escalate, in which fighting intensity grows in some way over the course of the conflict. Our analyses show that large wars are not unusually long wars, and large wars do not typically begin with intense fighting. Instead, conflict severities within civil and interstate wars show that escalation is the primary mechanism by which wars have become large over the past 200 years. In fact, high-variance escalation dynamics (large changes in fighting intensity) appear to be a generic feature of all forms of armed conflict (Fig.~2B). While most conflicts exhibit periods of relative stasis (little escalation or deescalation), each additional conflict year is associated with a 10\% \textit{ex ante} risk for increasing in severity by a factor of about 2, and a 1\% risk for increasing in severity by a factor of about 10 or more.

% 31. 
Hence, large wars become large because they escalate.  Research on conflict termination often emphasizes the role of strategic behavior, bargaining, and information gains, in which fighting provides information that allows the belligerents to update their assessments of relative forces and objectives~\cite{reiter:2010}. However, escalation dynamics indicates that these theories are incomplete. As fighting unfolds and uncertainty lowered, theories based on information gain predict that conflicts should tend to deescalate. Instead, we find that fighting is equally likely to escalate as it is to deescalate in the next year. This pattern may indicate a greater role for commitment problems in driving escalation dynamics within conflicts, and expanding theories of war termination to account for the role of severity is an important direction for future work.

% 32. 
Despite the generic pattern of escalation dynamics overall, civil wars appear different from interstate wars in one respect:\ very large civil wars exhibit a modest systematic tendency to deescalate (Fig.~2C). A simple model that combines the empirical variations in initial civil war sizes and their overall durations demonstrates that this slight tendency toward deescalation is sufficient to fully explain the observed distribution of civil war total sizes (Fig.~3B). Remarkably, this escalation model for civil wars is entirely nonparametric, with no fitted parameters, and yields a close agreement between simulated and historical civil war sizes since 1946. Robustness tests show that escalation dynamics are also necessary to explain civil war sizes, as a model without them fails to reproduce the largest civil wars (Fig.~3A).

% 33. 
A number of distinct factors or sociopolitical process may underlie the tendency for large civil wars to self-limit their size. For instance, because civil wars are defined as occurring within a single state, its total population may constrain the conflict's escalatory potential, by limiting the number of individuals who may potentially die in battle before the capacity to continue fighting erodes. Non-state belligerents may also have limited capacity to continue escalating compared to a state's greater capacity. Civil wars where the non-state actors develop greater capacity and can act like a state to some degree, e.g., as in the Chinese and American civil wars, offer an interesting test case for this hypothesis:\ we would expect those conflicts to have inherently more escalatory potential.

% 34.
Internationalization may also relax the tendency toward deescalation in large civil wars, making them behave more like interstate conflicts, e.g., if there is international support for the rebels. Examples of this kind of indirect intervention in the postwar era include the US Operation Cyclone to support the Mujahideen in Afghanistan 1979-1992 (with just below one million battle deaths over the period), or the extensive international support in the 1946-49 Civil War in Greece, which likely helped make it exceptionally large (154,000 battle deaths), for a limited population (7.4 million in 1946).

% 36. 
Escalation dynamics also explain the sizes of interstate wars, but only if we consider their unconditioned form, in which the risk of escalation is independent of conflict size. This difference implies that civil war sizes do not follow Richardson's Law for interstate war sizes~\cite{richardson:1948}. Instead, the deescalatory tendency of large civil wars, despite their longer durations, tends to suppress the occurrence of the very largest events. In contrast, interstate wars remain just as likely to further escalate, no matter how large the conflict has become so far. This behavior implies profound uncertainty for any risk assessment of ongoing or potential future conflicts, as we showed through two forecasting exercises (Fig.~5A-C). The First World War exemplifies this dynamic, where a pattern of alliances led to rapid escalation of hostilities across much of Europe following the assassination of Archduke Franz Ferdinand of Austria in Sarajevo in 1914.  Because of the connection between war size and cost, strategic considerations before or within an armed conflict that account for the potential for escalation may shed new light on war onset, termination, and other conflict outcome variables. For instance, in the ongoing war in Ukraine, the West's strategy of slowly increasing the level and sophistication of its aid may have prevented dramatic escalation by Russia.

% 38. 
An important limitation of our model is that we do not explain the initial sizes of armed conflicts when they first begin, even as initial size varies broadly and correlates poorly with the population of the involved state. For civil wars, at least, the largest countries are indeed where the largest conflicts tend to occur, but so too do the smallest. Understanding what factors predict this initial severity for either civil or interstate wars would substantially improve the utility of the escalation model in conflict forecasting tasks~\cite{cederman:weidmann:science:2017}. Similarly, our analysis relies on data that is aggregated at the yearly level, and hence the escalation factors we introduce represent an aggregation of many events and engagements that occur at finer time scales. Ideal conflict data would be comprehensive at the level of battles, which would allow an exploration of how escalation dynamics unfold at different levels of temporal aggregation.

% 40.
Characterizing how the conflict covariates shape the potential for escalation and deescalation within a conflict is an important direction for future work. For instance, are some types of conflict-level incompatibilities more likely to produce escalation within the conflict? A geographic incompatibility~\cite{cedermanetal:2015}, which a separatist group is able to operate in some ways like a state, may be inherently more risky for escalation than a military coup. Modeling escalation factors as a function of conflict-level covariates such as the type of incompatibility, geography, capabilities, characteristics of the belligerents, characteristics of allies, etc. would shed considerable light on precisely how escalation occurs.

% 41. 
Escalation dynamics may also have an endogenous component, in which escalatory potential depends on the particular way that the fighting unfolds or feedbacks between fighting and domestic politics. Untangling the relationship between within-conflict characteristics and within-conflict severity dynamics will likely require developing new theories and new data sets. For interstate wars in particular, it may be important to disaggregate at a finer timescale than yearly (as in Ref.~\cite{min:2021}), since these conflicts tend to have relatively short durations, and our simulation results suggest that escalation dynamics may run ``faster'' for interstate conflicts. 

% 42.
Escalation dynamics are a fundamental aspect of all forms of armed conflict, and our results indicate that they are both a necessary and sufficient explanation for how large wars become large. Previous theoretical models for the sizes of war have not accounted for the role of escalation within a conflict (see Ref.~\cite{cederman:2003}), and incorporating escalation will require developing new models of conflict duration and within-conflict severities. Escalation dynamics also imply a theoretical connection between conflict cost and war onset, due to the profound uncertainty they induce in just how large, and hence how costly, a conflict might become. Such uncertainty may serve as a deterrent to war onset, or may constrain certain actions within a conflict, in order to avoid escalation. Hence, the dynamics of escalation within a conflict offer a rich new window into armed conflict. A more fully developed theory of how escalation dynamics influence other, more well-studied aspects of international relations would be an important contribution to the field, and would guide efforts to understand how escalation dynamics depend on the characteristics of a conflict and its belligerents.

\bigskip
\noindent\textbf{Acknowledgements:}\ The authors thank H{\aa}vard Hegre, Bethany Lacina, Michael Colaresi, Bear Braumoeller, Nils Lid Hjort, Dale C.\ Copeland, Alexander Ray, and Daniel B.\ Larremore for helpful conversations, and thank the Centre for Advanced Study at the Norwegian Academy of Science for its support and in hosting AC and KSG while some of this work was conducted. A preliminary version of these results was first presented at the International Workshop on Coping with Crises in Complex Socio-Economic Systems at ETH Z\"urich in 2011. \smallskip \\
\textbf{Author Contributions:}\
conceptualization (AC, L-EC, KSG); data curation (AC, KSG); formal analysis (AC, KSG); methodology (AC, KSG); writing (AC, BFW, L-EC, KSG). \smallskip \\ 
\textbf{Competing Interest Statement:}\ The authors declare that they have no competing interests. \smallskip \\
\textbf{Data and materials availability:}\ All data needed to evaluate the conclusions in the paper are present in the paper and/or the Supplementary Materials. The PBD data used in this paper are available from the Peace Research Institute of Oslo at \url{https://www.prio.org/data/1/ }. The Correlates of War interstate war data (v4.0) used in this paper are available from the Correlates of War Project at \url{www.correlatesofwar.org/}.

%% BIBLIOGRAPHY

% macro to make Figures numbered as S#
\renewcommand{\thefigure}{S\arabic{figure}}
\setcounter{figure}{0}
% macro to make Tables numbered as S#
\renewcommand{\thetable}{S\arabic{table}}
\setcounter{table}{0}

%\newpage

\subsection*{Supplementary Material}

%% CONFLICT SIZE vs DURATION
\subsubsection*{Appendix A:\ Conflict severity and duration}
Historically, conflict severity $X$ (battle deaths) and conflict duration $T$ (years) are weakly or at best moderately correlated characteristics of civil and interstate wars (Fig.~S1), with substantial unexplained variance.

Pooling the 264 civil and internationalized civil wars 1946--2008 in the PRIO Battle Deaths data and the 95 interstate wars 1823--2003 in the Correlates of War data, severity and duration are weakly correlated ($r^2 = 0.15$, Pearson correlation, $\log X$ vs.\ $T$, $p < 10^{-14}$).

However, some of this lack of correlation is caused by the different sizes and durations of civil and interstate conflicts. For civil and internationalized civil wars alone, severity and duration correlate at the $r^2 = 0.30$ level (Pearson correlation, $\log X$ vs.\ $T$, $p < 10^{-20}$), indicating that the largest civil wars also tend to have long durations. At the same time, however, many of the largest civil wars also had very short durations. What are not observed are long conflicts with very low total severities. And, for interstate wars alone, severity and duration correlate at the $r^{2}=0.45$ level (Pearson correlation, $\log X$ vs.\ $T$, $p < 10^{-10}$), indicating at best a moderate correlation between size and duration. What tend not to be observed are short but very large interstate wars or long but not very large interstate wars.

%% FIGURE S1
\begin{figure}[h!]
\begin{center}
\includegraphics[scale=0.39]{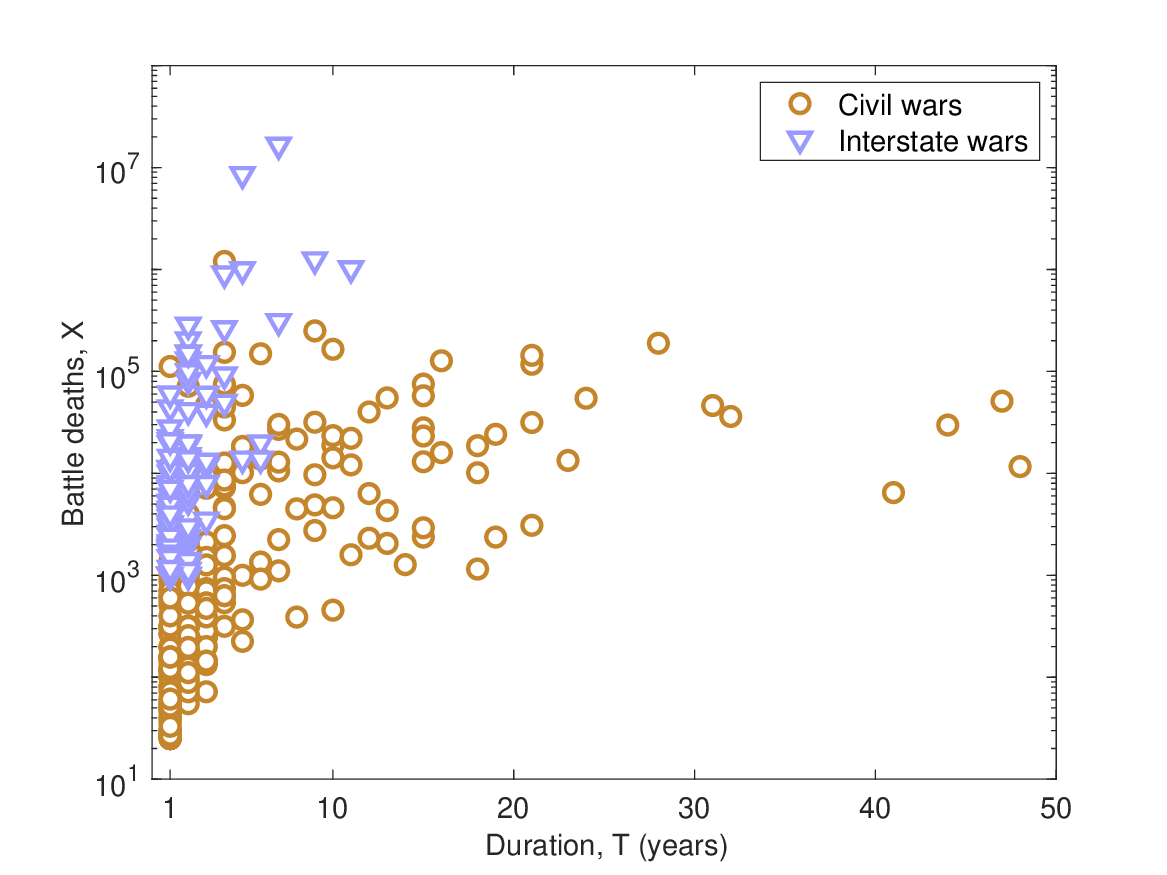}
\end{center}
\caption{ Conflict total severity (battle deaths) as a function of conflict duration (years), for 264 civil and internationalized civil wars 1946--2008 in the PBD and 95 interstate wars 1823--2003 in the CoW, showing weak correlation between severity and duration.}
\label{fig:S1}
\end{figure}
% ---

\subsubsection*{Appendix B:\ Civil and interstate war data}
Most existing work on war sizes focuses on all wars, both civil and interstate, or considers only data for interstate wars~\cite{braumoeller:2019,cirillo:taleb:2016,clauset:2018,richardson:1948}. Richardson's foundational work~\cite{richardson:1948} argued that the frequency-severity distribution of conflict severity $X$ follows a power-law distribution of the form $\Pr(X) \propto X^{-\alpha}$, where $\alpha>1$ is a parameter. Richardson's original data on ``deadly quarrels'' combined interstate wars, civil wars, and what are now often called intercommunal conflicts between non-state actors not involving the state. This analytical approach implicitly assumes that all armed conflict stems from a common data generating mechanism, irrespective of the type of actors involved (i.e., states and non-state actors) or other factors, and neglects their categorical difference---civil wars unfold within one state, while interstate wars unfold among two or more. This assumption is debated in the literature, and it remains an open question as to what aspects of different types of conflict are common and what aspects are distinct.

The PRIO Battle Deaths (PBD) data is based on the definitions of the Uppsala Armed Conflict Data (ACD), which identifies conflicts on the basis of ``a contested incompatibility that concerns government and/or territory, where \ldots armed force \ldots results in at least 25 battle-related deaths'' in a calendar year \citep[618-9]{gledetal:2002}, and covers all such conflicts worldwide 1946--2008. Distinct conflict identifier codes are assigned based on the specific incompatibilities (territory or government of a country). 

For each conflict-year, the PBD reports a high, low, and ``best'' estimate. In our analyses, we strictly impose the ACD definition, and discard records with a best annual estimate $x_{t}<25$; this criterion resulted in discarding 3~years of civil war data. For the remaining cases, best estimates for battle deaths in a given year that were coded as missing were replaced with the geometric mean of the recorded low and high estimates, which reflects the analytic frame of proportional changes that we use in our statistical analyses. Some conflicts in the PBD have ``gaps,'' i.e., years without violence above the 25 deaths threshold between active conflict periods. We follow the standard approach in research conflict duration to ignore gaps of less than two calendar years~\cite{wmcg:2012}; if a conflict contains one or more gaps that exceed this threshold, we break it into contiguous sequences that themselves satisfy the gap condition.

The resulting data set is disaggregated below the level of entire conflicts, and aggregated at the annual level. That is, the annual severity $x_{t}$ remains an aggregated measure of conflict dynamics, and hence should not be interpreted as a measure of average conflict intensity at smaller time scales, e.g., at the level of weeks or days, or across space.

The data include conflicts that were ongoing during 2008. For these conflicts, we cannot estimate the final total duration $T$ or total severity $X$, but we do include their within-conflict escalation factors $\lambda_{t}$ in our analysis as these are uneffected by the right-censoring. There are also some left-censored conflicts with onsets prior to 1946, e.g., the Chinese civil war. Moreover, conflicts in some countries such as Angola are arguably left-censored, since they grow out of colonial conflicts prior to independence. We identify conflicts strictly based on the assigned identifiers, and do not try to make independent judgements on whether something should be coded as the ``same'' or a distinct conflict in the same countries.

A common step in analyzing conflict severities is to normalize the severity by the country's population in order to obtain a conflict ``intensity'' value. Such a normalization makes the implicit assumption~\cite{cranmer:desmarais:2016} that all individuals in the country are at equal risk of dying in the conflict, which is not generally true. A more accurate measure of intensity would require a model that estimates the effective population as risk of dying, but such models are not generally available. In our study, the escalation factors implicitly account for the country's population because they are derived from the ratio of two severities (see Appendix~C); hence, so long as population change year-by-year is relatively small, an escalation factor derived from absolute severities will closely approximate an escalation factor derived from intensities.

\subsubsection*{Appendix C:\ Additional analysis and simulation results}
\paragraph{Civil and interstate war analyses.}

Although civil and internationalized civil wars exhibit a systematic tendency to deescalate when their annual severity is above $x_t>500$ (Fig.~2C), we find no comparable tendency among interstate conflicts 1946--2008 (Fig.~S2A). Moreover, the tendency for large civil wars to deescalate is fairly smooth in the $x_t>500$ range, such that as annual severity $x_t$ increases, the escalation factor becomes progressively more deescalatory, roughly in proportion to the log of severity $\log x_t$. In contrast, the mean escalation factor for interstate wars remains very close to $\lambda=1$ in the $x_t>500$ range, indicating no systematic tendency to escalate or deescalate.

In contrast, interstate and civil wars both exhibit a systematic tendency to escalate when their annual severity is small, and this pattern closely mirrors the same pattern observed in civil wars. In the PBD, both civil and interstate wars have a minimum annual severity $x_{\textrm{min}}=25$, and so this effect is attributable to the left-censoring effect here:\ conditioned on the conflict continuing into the $t+1$ year, the distribution of escalation factors is truncated below $\lambda_{\textrm{min}} = x_{\textrm{min}} / x_t$, which has the effect of shifting the mean upward, above 1. As a result, we do not regard this systematic tendency at the low end to be nearly as interesting as the presence or absence of a systematic deviation from $\lambda=1$ at the upper end.

%% FIGURE S2
\begin{figure}[h!]
\begin{center}
\includegraphics[scale=0.39]{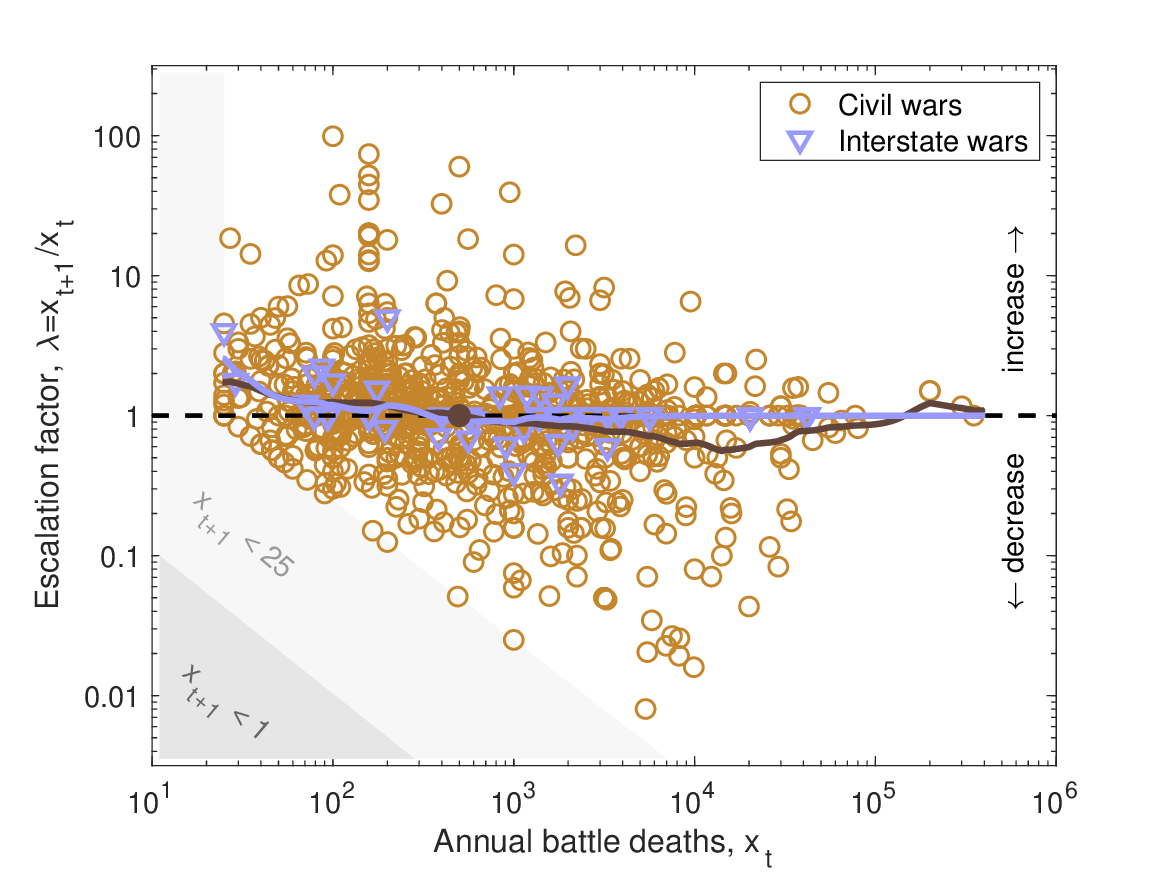} \\
\includegraphics[scale=0.39]{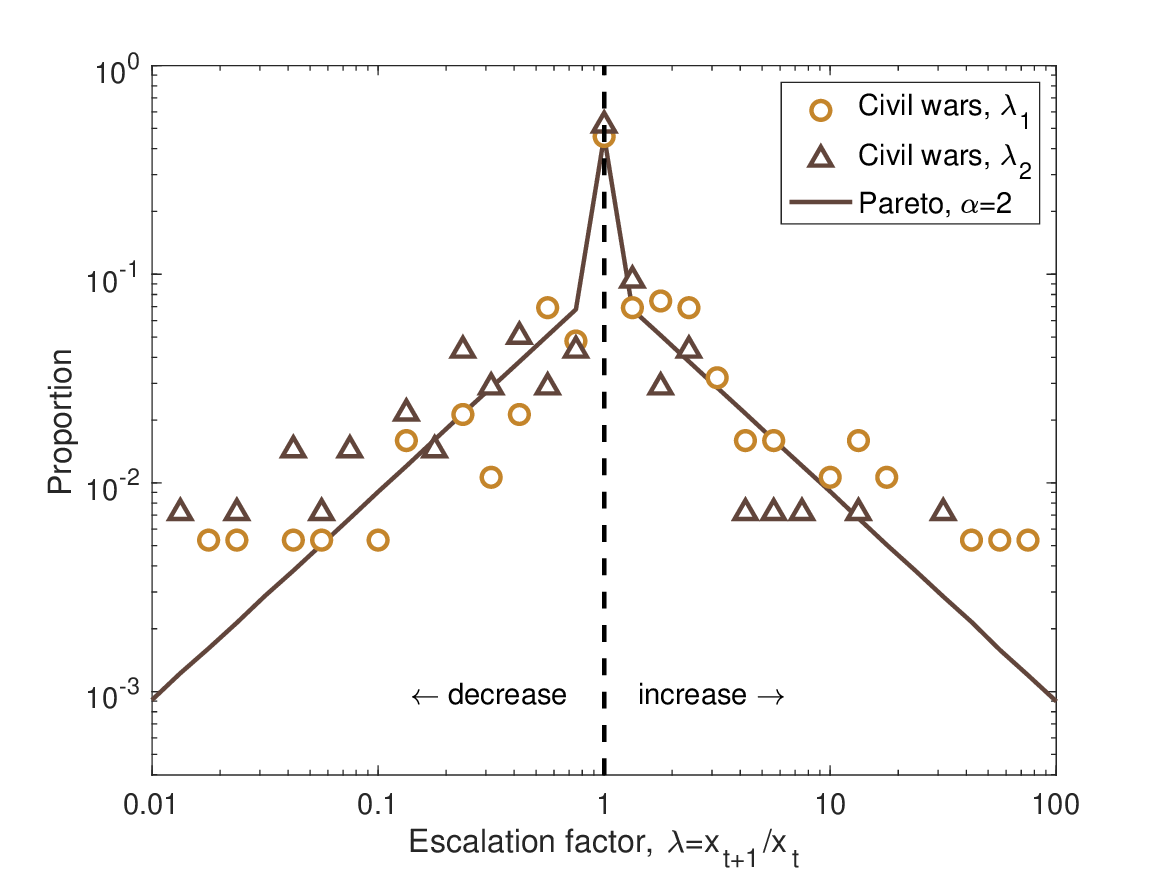}
\end{center}
\caption{ (A) Joint distribution $\Pr(\lambda \,|\, x_{t})$ of escalation factors and annual severity $x_{t}$ for civil wars (repeated from Fig.~2C) 
and for interstate wars contained in the PBD. For both types of conflicts, the solid line indicates the smoothed geometric mean pattern, indicating that in both cases, small conflicts tend to escalate in severity ($\langle \lambda \rangle >1$). However, among large conflicts, only civil wars exhibit a systematic tendency to deescalate, while interstate wars show no systematic tendency in either direction. (B) Distributions of first $\lambda_1$ and second $\lambda_2$ escalation factors for civil and internationalized civil wars in the PBD, showing a slight tendency for $\lambda_2$ to be deescalatory ($\langle \lambda \rangle <1$). }
\label{fig:S2}
\end{figure}
% ---

The distribution of annual escalation factors $\Pr( \lambda_t )$ among conflicts that last at least $t+1$ years, provide some insight into the question of whether escalation factors can be treated as iid random variables. Overall, we find that the first and second escalation factor distributions $\Pr( \lambda_1 )$ and $\Pr( \lambda_2 )$ are very similar to the overall distribution of escalation factors, exhibiting high variance forms with roughly Pareto distribution tails (Fig~S2B). However, the first factor tends to be escalatory $\langle \lambda_1 \rangle = 1.25$ (geometric mean) while the second factors tends to be deescalatory $\langle \lambda_1 \rangle = 0.85$.
\\

\paragraph{The escalation model.}
Generating a simulated conflict time series $x_1,x_2,\dots,x_T$ requires three choices. First, the initial severity in the first year of fighting $x_1$ must be chosen. Second, the length of the time series $T$ (years) must be chosen. Finally, for each year $2\leq t < T$, each annual severity is given by the update equation
\begin{align}
x_{t+1} & = \lambda_t \, x_t \enspace ,
\label{eq:S1}
\end{align}
where $\lambda_{t}$ is the escalation factor that quantifies how the fighting intensity changes from year $t$ to year $t+1$. In this way, the annual severity $x_{t}=x_{1}\,\lambda_{1}\,\lambda_{2}\dots \,\lambda_{t-1}$, where the series of $\lambda$ factors denotes the cumulative escalation and deescalation effects over the conflict's history up to that point. Mathematically, the total severity is then given by
\begin{align}
X&= \sum_{t=1}^{T}x_{t} \nonumber \\
&= x_{1}\!\left[\sum_{t=1}^{T}\lambda_{1}\,\lambda_{2}\dots \,\lambda_{t-1}\right] \nonumber \\
&=x_{1}\!\left[\sum_{t=1}^{T}\left(\prod_{\tau=1}^{t-1}\lambda_{\tau}\right)\right] \enspace .
\end{align}
The leading position of the initial severity in this expression illustrates its fundamental role, effectively setting the scale of the full conflict.

%% FIGURE S3
\begin{figure*}[t!]
\begin{center}
\begin{tabular}{ccc}
\includegraphics[scale=0.30]{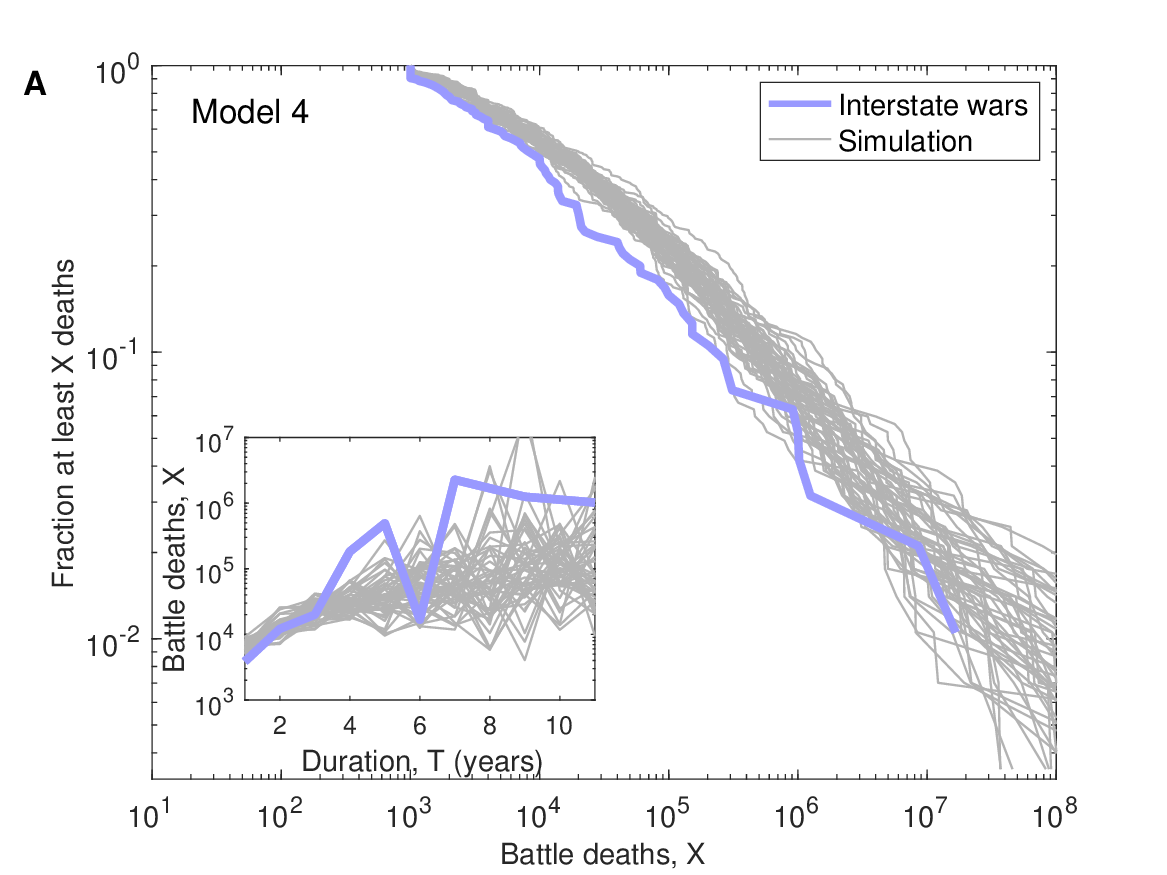} &
\includegraphics[scale=0.30]{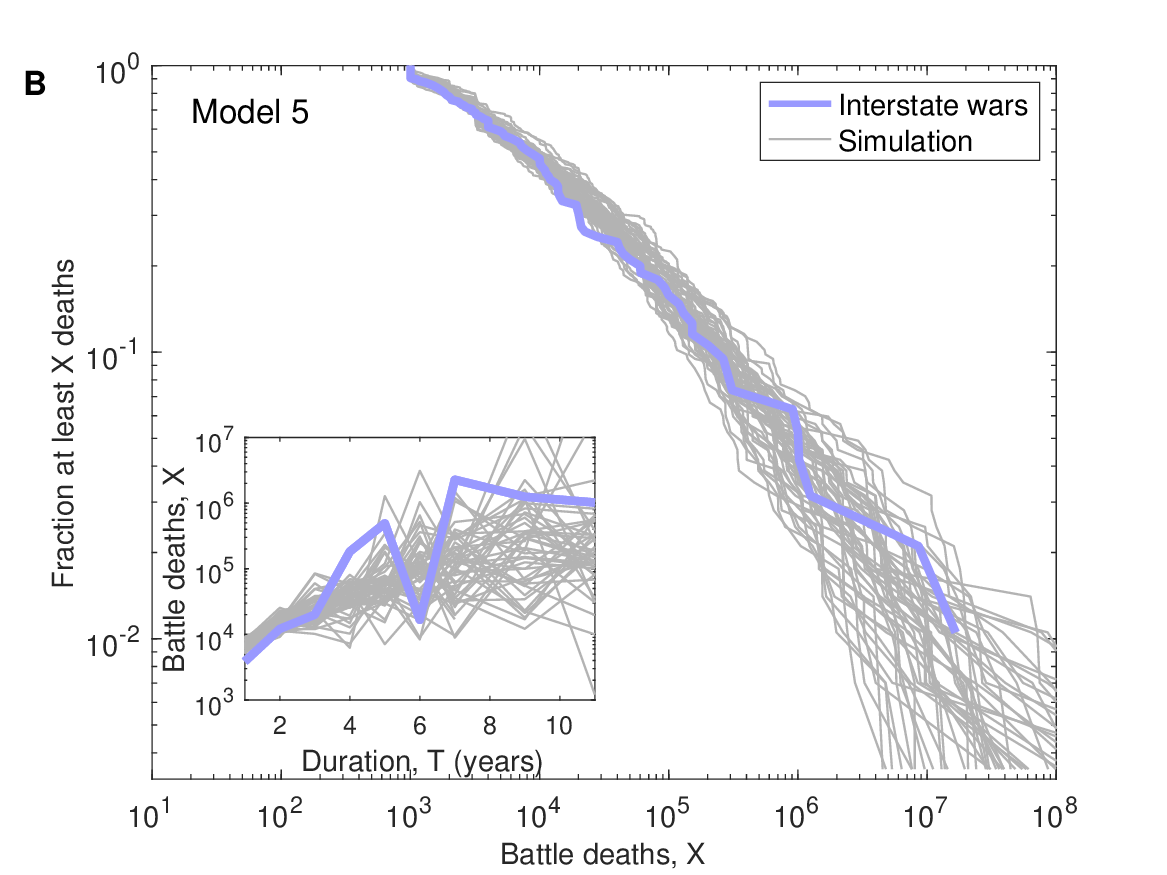} &
\includegraphics[scale=0.30]{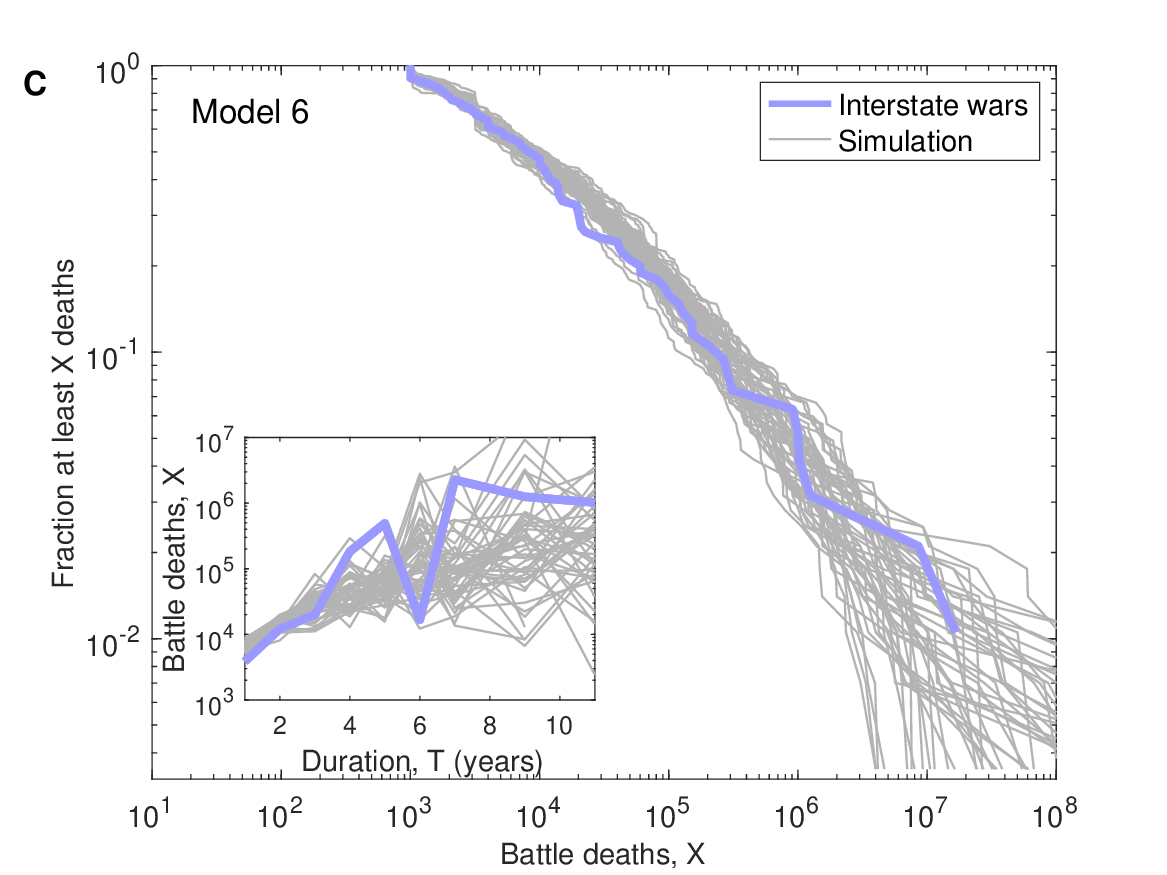} \\
\end{tabular}
\end{center}
\caption{\textbf{Simulating interstate war severities.} Three additional variations on the escalation model for interstate war sizes, where (A) Model~4 (long wars) draws conflict duration $T$ from the empirical distribution $\Pr(T)$ for civil wars, (B) Model~5 (more amplification) draws three factors from $\Pr(\lambda)$ for each time step, and (C) Model~6 (db Pareto) draws each factor from a parametrized double Pareto distribution that has the same tail structure as the empirical distribution but eliminates the artifact at $\lambda=1$. Insets:\ the mean conflict severity $\langle X \rangle$ as a function of conflict duration $T$ (years) for interstate wars and the corresponding Model simulations. } 
\label{fig:S3}
\end{figure*}
% ---

In this model of escalation dynamics, the annual severity $x_{t}$ evolves as a multiplicative random walk according to Eq.~(1), and the total severity $X$ is the discrete integral over its trajectory. A key feature of this random walk model is that the length of the walk $T$ is itself a random variable. Hence, the distribution of total severities $\Pr(X)$ is related to but distinct from the expected log-normal distribution that often appears in asymptotic analyses of multiplicative random walks.

Its simplicity makes simulating the escalation model straightforward. Given a distribution of initial conflict severities $\Pr(x_{1})$, a distribution of durations $\Pr(T)$, and a model for choosing escalation factors $\lambda$, we first initialize the simulation by drawing $x_{1}$ and $T$, and then iterate the update Eq.~(1), drawing a new factor $\lambda$ each time. Here, we make this model fully non-parametric by using the empirical distributions of initial severities, durations, and escalation factors for each of these three parts. Parametric distributions could be used instead, but we do not explore these variations here, except in one case. Similarly, these distributions could be replaced with models that depend on conflict covariates, which would be a valuable direction for future work.

For the civil war escalation model only, in order to both capture the empirical structure in the joint distribution $\Pr(\lambda\,|\,x_{t})$ and provide a way to map simulated annual severities to a distribution of escalation factors, we use a coarsening approach. Specifically, we bin the empirical annual severities $x_{t}$ by order of magnitude and then tabulate a set of four conditional distributions on the associated escalation factors. The bin boundaries are chosen to be $[25, 100]$, $(100,1000]$, $(1000,10000]$, and $(10000,\infty)$; we obtain similar results for similar logarithmic binning schemes. Then, given a simulated annual severity $x_{t}$, we identify the corresponding bin based on its size, and then draw an escalation factor iid from that bin's distribution of factors. This approach produces a non-parametric form of $\Pr(\lambda\,|\,x_{t})$ that can be used in the simulation despite the fact that most simulated values of $x_{t}$ will not appear precisely in the empirical data.
\\

\paragraph{Interstate war simulations.}
We consider six variations on the escalation model to explain the sizes of interstate wars. In each model, we enforce the same criteria as were used in the PBD for recording conflict severities. Specifically, for each annual severity $x_{t}$, we enforce a minimum value of $x_{\textrm{min}}=25$, and for the total conflict severity~$X$, we enforce a minimum value of $X_{\textrm{min}}=1000$. If a simulated severity is below its respective minimum, we replace the out-of-bounds value with that minimum.

In all models, unless otherwise noted, the initial size $x_{1}$ is a constant factor of $20$ times an iid draw from the empirical distribution $\Pr(x_{1})$ for initial severities in civil wars 1946--2008, and the conflict's duration $T$ is drawn iid from $\Pr(T)$ for interstate wars 1823--2003.

\begin{itemize}
\item \textbf{Model~1} (no escalation):\ Total severity is a function of initial severity and duration only, i.e., \mbox{$X = x_{1}\times T$}. 

Initial severity $x_{1}$ is drawn iid from $\Pr(X\,|\,T\!=\!1)$, the set of total severities for interstate wars 1823--2003 that lasted $T\!=\!1$ year. 
\item \textbf{Model~2} (escalation):\ Each factor $\lambda$ is drawn iid from the unconditioned distribution of escalation factors $\Pr(\lambda)$ (Fig.~2B).
\item \textbf{Model~3} (amplified escalation):\ A factor $\lambda$ is drawn iid from $\Pr(\lambda)$, and then squared $\lambda^{2}$.
\item \textbf{Model~4}  (long wars):\ Duration $T$ is drawn iid from $\Pr(T)$ for civil wars, and a factor $\lambda$ is drawn iid from $\Pr(\lambda)$.
\item \textbf{Model~5}  (more amplification):\ Three factors $\lambda_{1}$, $\lambda_{2}$, $\lambda_{3}$ are drawn iid from $\Pr(\lambda)$, and then multiplied.
\item \textbf{Model~6} (db Pareto):\ Each factor $\lambda$ is drawn iid from a double Pareto distribution with parameter $\alpha=2$, which lacks the artifact at $\lambda=1$ seen in the empirical distribution.
\end{itemize}

Simulation results compare 50 repetitions of a particular model with the empirical interstate war size distribution $\Pr(X)$ and the empirical size $X$ vs.\ duration $T$ function. This approach illustrates the variability of each model's output, and the empirical distribution provides a reference point against which to identify meaningful deviations between model and data. Results for Models~1,~2, and~3 are given in the main text in Figure~4A-C, and results for Models~4,~5, and~6 are shown in Figure~S3A-C.

Model~4 is the same as Model~1 (escalation) but with conflict durations drawn from the distribution of civil war durations, which are substantially longer than interstate wars. Hence, this model produces simulated interstate conflicts that are systematically larger than those observed historically, suggesting that the typical short duration of interstate conflicts does tend to suppress the total size of interstate wars but it does not change the shape of the distribution's tail.

Model~5 is the same as Model~3 (amplified escalation) but with an additional escalation factor (three vs.\ two). Model~6 is the same as Model~1 (escalation) but replaces the empirical distribution of escalation factors with a double Pareto distribution with a tail parameter $\alpha$ estimated from the empirical data. This variation smooths out the artifact in the empirical $\Pr(\lambda)$ at $\lambda=1$ and hence tests whether that feature is important to explain the sizes of wars. Both Models~5 and~6 produces good agreement with the historical variation in interstate war sizes, and the scaling relationship between size and duration, at a similar level as Model~3.

The key takeaway from this modeling exercise is that all of the models produce size distributions that are roughly similar to empirical data, albeit with some differences. This commonality indicates that the variations among the models are less important than their common features of (1)~an unconditioned distribution of escalation factors $\Pr(\lambda)$, which allows large conflicts to become even larger (in contrast to the size-conditioned distribution for civil wars), and (2)~a scaling up of the initial size $x_{1}$, which captures the overall greater intensity of fighting in interstate wars as compared to civil wars.

\subsubsection*{Appendix D:\ Forecasting civil war}

For a hypothetical civil war in a particular country, we draw the initial severity $x_{1}$ from a model that relates the initial conflict intensity $\gamma$ to the population in that country in the year of conflict onset. Empirically, the initial intensity of a conflict is given by the ratio \mbox{$\gamma=x_{1}/p_{1}$}, where $x_{1}$ is the number of battle deaths in year $t=1$ of the conflict and $p_{1}$ is the population of that country in that same calendar year.

%% FIGURE S4
\begin{figure}[t!]
\begin{center}
\includegraphics[scale=0.39]{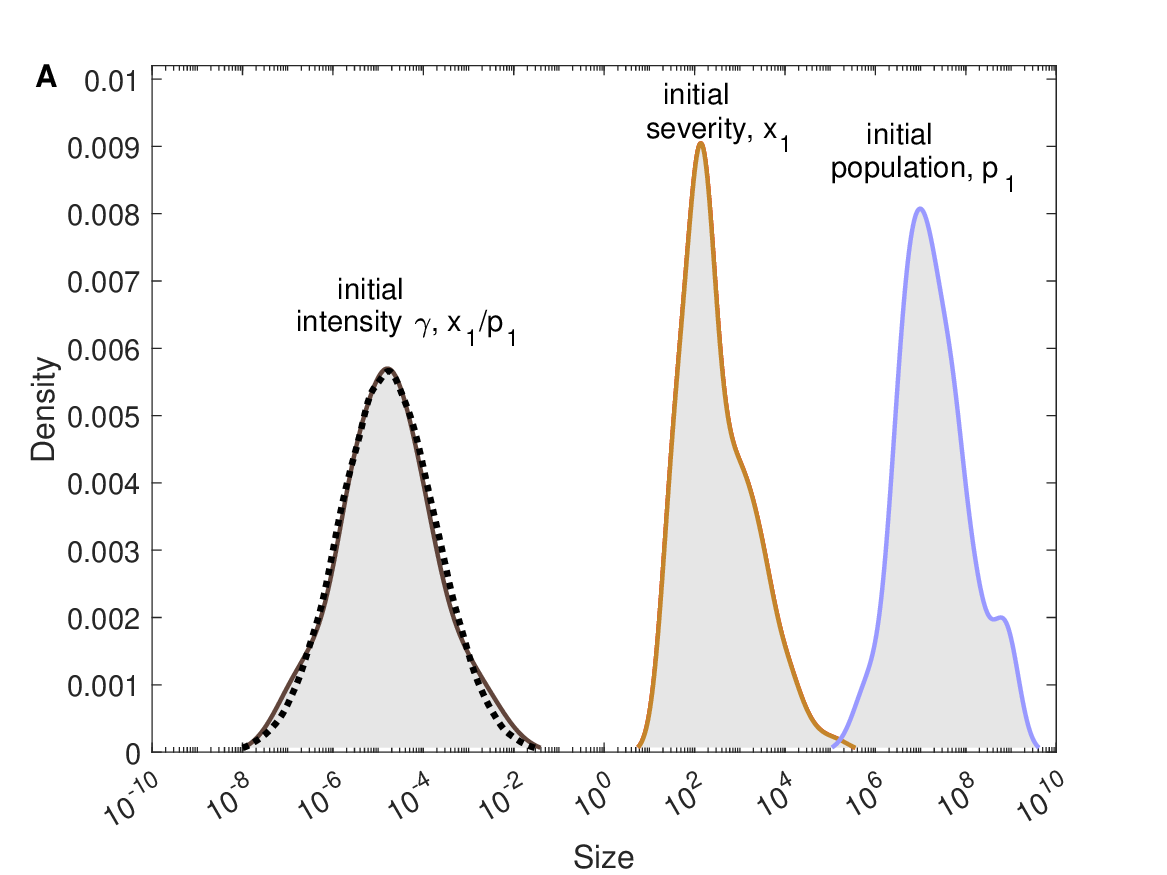} \\
\includegraphics[scale=0.39]{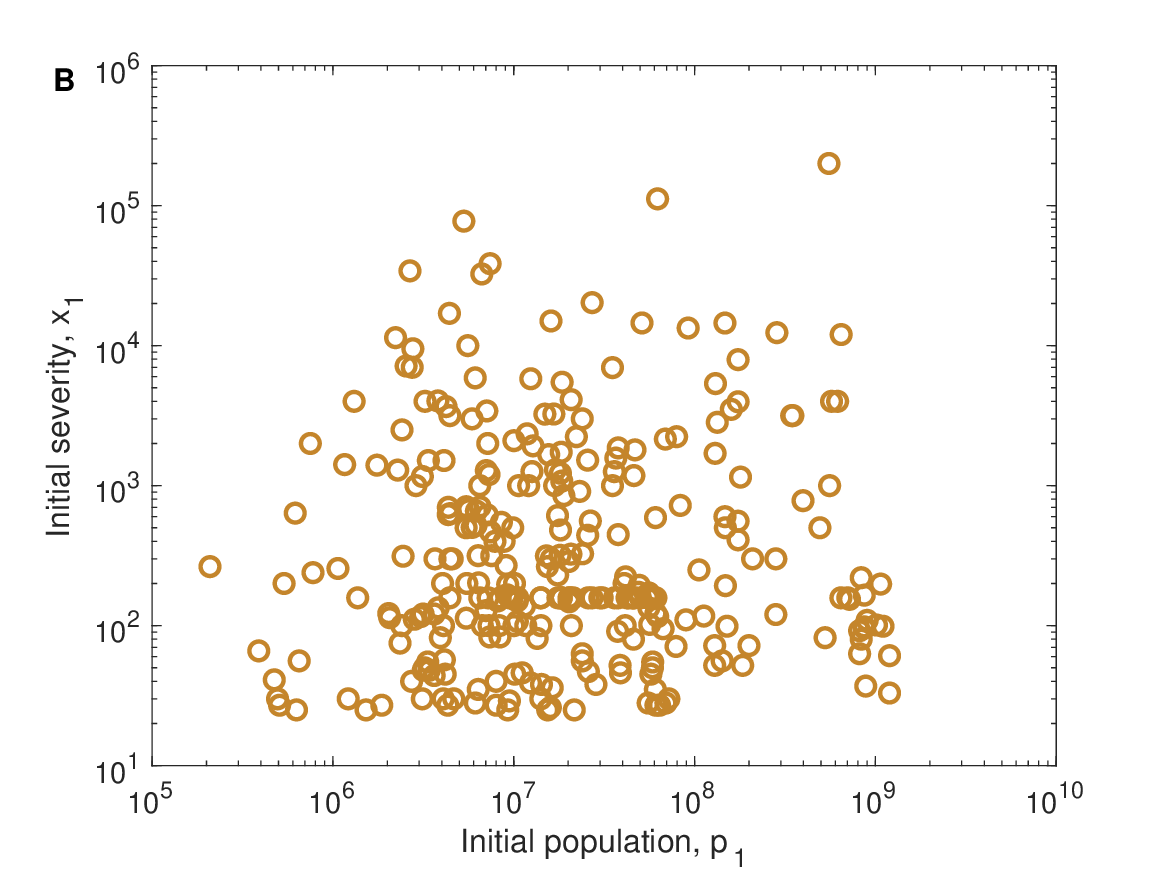}
\end{center}
\caption{\textbf{Initial severity of civil wars.} (A) Empirical distributions (smoothed) for initial severity $\Pr(x_{1})$ and initial population $\Pr(p_{1})$, along with the corresponding distribution of conflict initial intensities $\Pr(\gamma)$ fitted with a log-normal distribution $\mu = 1.89\times10^{-5}$ and $\sigma =   2.45$. Mean initial severity (geometric) is $315$ battle deaths, with a mode of 141. (B) Initial severity $x_{1}$ versus initial population $p_{1}$, showing no significant correlation ($r^{2}<0.01$, $p=0.63$). }
\label{fig:S4}
\end{figure}
% ---

We construct the empirical distribution of intensity factors $\Pr(\gamma)$ by matching the country's population according to U.N.\ population estimates~\cite{unpopdiv} in each year $t$ of a given conflict time series with the corresponding annual severity in that year. The distribution of initial populations and the distribution of initial severities are very roughly log-normally distributed, and the distribution of intensities $\Pr(\gamma)$ is slightly broader than both (Fig.~S4A), with the modal intensity being $\gamma=1.89\times10^{-5}$ of the initial population dying in the first year of conflict. However, the variance in these intensity factors is large, with the smallest intensity being close to $\gamma=10^{-8}$ and the largest being around a staggering $\gamma=10^{-2}$. Moreover, initial severity and initial population are not significantly correlated (Pearson correlation, $\log x_{1}$ vs.\ $\log p_{1}$, $r^{2}<0.01$, $p=0.63$), in large part because both small and large conflicts occur in large countries (Fig.~S4B).

For the forecasting tasks with hypothetical conflicts, we draw factor $\gamma$ uniformly at random from this empirical distribution and multiply it by the country's estimated  population in that year; each replication of the forecasting model draws a new intensity factor, which drives substantial variability in the initial size of the conflict.

For an ongoing civil war, we use the last observed annual severity in 2008 to seed the escalation model as $x_{0}$, drawing an escalation factor from the joint distribution $\Pr(\lambda \,|\,x_{t})$ in order to set $x_{1}$, the first annual severity in the simulated time series. 

Civil war durations are substantially longer, on average, than interstate war durations. To simulate the evolution of an in-progress civil war, we assume that a conflict's duration is governed by a coin-flipping process, in which the probability that an ongoing conflict stops in year $t$ is given by a hazard function $\Pr(\rm{stop}\,|\,t)$. If the hazard function $\Pr(\rm{stop}\,|\,t)=$ const., then stopping is an iid process and durations $T$ follow a geometric (discrete exponential) distribution.

We consider two parametric forms of civil war durations $\Pr(T)$---a discrete Weibull (stretched exponential) distribution and a piecewise geometric distribution with a breakpoint at $t=5$ years---which we fit to the empirical duration data for civil conflicts 1946--2008. We exclude the 35 conflicts that meet the definition of ``ongoing'' from this calculation, leaving 264 conflict durations. The probability density of the discrete Weibull has the form
\begin{align}
\Pr(T) & \propto T^{\beta-1} {\rm e}^{-\lambda\,T^{\beta}} \enspace , \nonumber
\end{align}
where $\beta$ and $\lambda$ are the parameters, and the probability density of the piecewise geometric has the form
\begin{displaymath}
\Pr(T) \propto \left\{ 
\begin{array}{ll}
{\rm e}^{-\lambda_{1}\,T} & \textrm{if $T< 5$} \\
{\rm e}^{-\lambda_{2}\,T} & {\rm otherwise , } \\
\end{array}
\right.
\end{displaymath}
where $\lambda_{1}$ and $\lambda_{2}$ are the parameters.

We note that in the special case of $\beta=1$, the discrete Weibull reduces to a geometric distribution with parameter $\lambda$; when $\beta<1$, the distribution is heavier-tailed than a geometric distribution, implying that the hazard rate for stopping $\Pr(\rm{stop}\,|\,t)$ is a decreasing function, i.e., the probability of stopping becomes smaller as a conflict duration becomes longer. Similarly for the piecewise geometric distribution, when $\lambda_{1}=\lambda_{2}$, the distribution reduces to a single geometric distribution; when $\lambda_{2}<\lambda_{1}$, the tail of the distribution decays more slowly than the body, implying that the hazard rate for stopping $\Pr(\rm{stop}\,|\,t\geq 5) < \Pr(\rm{stop}\,|\,t<5)$, i.e., the probability of stopping is smaller (governed by $\lambda_{2}$) if the conflict duration lasts more than 4 years.

Maximum likelihood estimates (MLE) of these model parameters are $\hat{\beta}=0.36\pm 0.05$, $\hat{\lambda}=1.35\pm0.34$ for the Weibull, and $\hat{\lambda}_{1}=0.57\pm0.05$, $\hat{\lambda}_{2}=0.09\pm0.01$ for the piecewise geometric. Figure~S5A shows the empirical duration distribution $\Pr(T)$ along with these maximum likelihood fits and the estimated hazard functions. (Uncertainty in parameter estimates denotes the standard deviation of the bootstrap distribution of maximum likelihood parameters; Figure~S5B,C.) In both cases, the hazard rate is convex, and most of the dynamic range occurs at short durations. That is, under a coin-flipping model for stopping, the hazard rate falls rapidly over the first few years of a conflict and conflicts that continue past this initial phase tend to have very low probabilities of stopping each year.

In our forecasting exercise, we count the preceding years of conflict in determining the simulation's hazard rate for termination. Because of the overall similarity of the two models, we use the simpler, piecewise geometric model for the forecasted conflict time series. We note that all of the selected ongoing conflicts had already continued for more than 4 years, and hence the hazard rate for stopping was $\hat{\lambda_{2}}$.

%% FIGURE S2
\begin{figure}[t!]
\begin{center}
\includegraphics[scale=0.38]{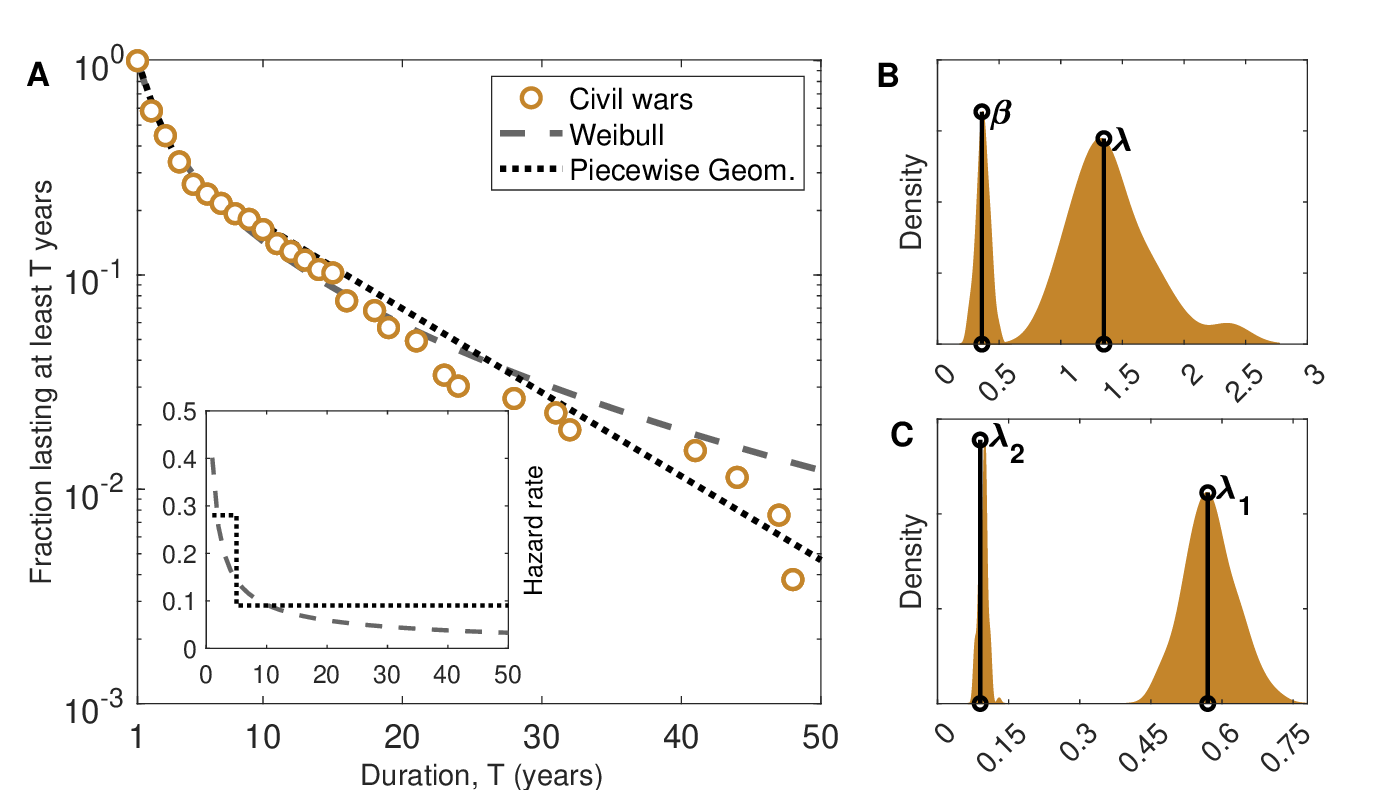}
\end{center}
\caption{\textbf{Modeling civil war durations.} (A)~Empirical distribution of conflict durations (years) for civil wars 1946--2008, along with Weibull and piecewise exponential fitted distributions.\ (inset) Hazard functions showing the modeled probability that a conflict with current duration of $t$ years will stop in that year, for the two fitted models. (B,C) Bootstrap distributions of maximum likelihood parameters, for Weibull and piecewise geometric distributions, respectively, along with MLE values from the empirical data (black lines). }
\label{fig:S5}
\end{figure}
% ---

\end{document}